\documentclass[lettersize,journal]{IEEEtran}
\usepackage{amsmath}
\usepackage{graphicx}
\usepackage{caption}
\usepackage{amsfonts}
\usepackage{color}
\usepackage{subcaption}
\usepackage{amsthm}
\usepackage{amssymb}

\newtheorem{theorem}{Theorem}

\IEEEoverridecommandlockouts
\begin{document}
\title{Delay-Doppler Pulse Shaping in Zak-OTFS Using Hermite Basis Functions}
\author{Fathima Jesbin and Ananthanarayanan Chockalingam\thanks{The authors are with the Department of ECE, Indian Institute of Science, Bangalore 560012, India. Email: \{fathimaj,achockal\}@iisc.ac.in.} 
}
\maketitle

\begin{abstract} 
The performance of Zak-OTFS modulation is critically dependent on the choice of the delay-Doppler (DD) domain pulse shaping filter. The design of pulses for $L^2(\mathbb{R})$ is constrained by the Balian-Low Theorem, which imposes an inescapable trade-off between time-frequency localization and orthogonality for spectrally efficient systems. In Zak-OTFS, this trade-off requires balancing the need for localization for input/output (I/O) relation estimation with the need for orthogonality for reliable data detection when operating without time or bandwidth expansion. The well-known sinc and Gaussian pulse shapes represent the canonical extremes of this trade-off, while composite constructions such as the Gaussian-sinc (GS) pulse shape offer a good compromise. In this work, we propose a systematic DD pulse design framework for Zak-OTFS that expresses the pulse as a linear combination of Hermite basis functions. We obtain the optimal coefficients for the Hermite basis functions that minimize the inter-symbol interference (ISI) energy at the DD sampling points by solving a constrained optimization problem via singular value decomposition. For the proposed class of Hermite pulses, we derive closed-form expressions for the I/O relation and noise covariance in Zak-OTFS. Simulation results of Zak-OTFS with embedded pilot and model-free I/O relation estimation in Vehicular-A channels with fractional DDs demonstrate that the optimized pulse shape achieves a bit error rate performance that is significantly superior compared to those of the canonical sinc and Gaussian pulses and is on par with that of the state-of-the-art GS pulse, validating the proposed framework which provides greater design flexibility in terms of control of ISI and sidelobe energies.
\end{abstract}
\vspace{1mm}

\section{Introduction}
\label{sec:intro}
Orthogonal time frequency space (OTFS) modulation is a delay-Doppler (DD) domain modulation designed to provide robust performance in wireless channels characterized by high Doppler spreads \cite{otfs1}-\cite{ssdas_book}. While early works on OTFS focused on the multicarrier version of OTFS \cite{otfs1}-\cite{h_b_mishra}, Zak transform based OTFS (Zak-OTFS) has emerged \cite{zak_otfs1},\cite{zak_otfs2} as a compelling alternative, offering a rigorous mathematical framework and enhanced robustness over a wider range of delay and Doppler spreads \cite{zak_otfs3}-\cite{sm2}. In Zak-OTFS, the information symbols are multiplexed in the DD domain, and the conversion to a time domain signal for transmission is done using inverse Zak transform. At the receiver, the received time domain signal is converted to DD domain using Zak transform. 

A critical component of the Zak-OTFS transceiver is the DD domain transmit pulse shaping filter, which is used to limit bandwidth and time duration of transmission. The choice of this filter is not trivial as it directly shapes the effective channel, which is a cascade of the transmit filter, the physical channel, and the receive filter. Consequently, the characteristics of the pulse in terms of localization and orthogonality in the DD domain profoundly influence the performance of the two primary receiver tasks, namely, estimation of the DD input-output (I/O) relation and subsequent equalization and detection of data symbols. The pulse shaping filter, therefore, should be carefully designed to enable both of these tasks to be performed with good accuracy and efficiency.

The design of pulses for the space of square-integrable functions on $\mathbb{R}$, $L^2(\mathbb{R})$, is constrained by the Balian-Low Theorem, which imposes an inescapable trade-off between time-frequency localization and orthogonality for spectrally efficient systems \cite{blt}. Good localization, characterized by concentrated energy and low side lobes, is essential for accurate I/O relation estimation in Zak-OTFS, as it reduces DD aliasing (i.e., interference from quasi-periodic replicas) and pilot-data interference. Conversely,
orthogonality which manifests as nulls at the DD domain Nyquist sampling points, is critical for minimizing inter-symbol interference (ISI) energy which is beneficial for reliable data detection. The well-known sinc and Gaussian pulse shapes represent the canonical extremes of this trade-off. The sinc pulse has perfect orthogonality (nulls at sampling points) by satisfying the Nyquist criterion for zero ISI but poor localization (high sidelobes), whereas the Gaussian pulse achieves 
optimal localization (very low sidelobes) by reaching the minimum time-bandwidth product allowed by the Heisenberg uncertainty principle at the expense of significant ISI (no nulls at sampling points). Root-raised cosine (RRC) pulse gives lower sidelobes compared to sinc pulse, but at the cost of bandwidth/time expansion. Composite pulse designs, such as the Gaussian-sinc (GS) pulse which inherits the complementary strengths of Gaussian and sinc pulses, offer a better compromise \cite{gs}. However, sinc, Gaussian, and GS pulses do not offer design flexibility without bandwidth or time expansion.

In this paper, we move beyond the well-known sinc, Gaussian, and GS pulses, and propose a systematic pulse design framework for Zak-OTFS with no bandwidth or time expansion. The core of our framework is to express the pulse as a linear combination of Hermite basis functions.
This approach has a strong foundation in multicarrier systems \cite{mc_haas_tfl},\cite{ofdm_tfl}, where it has been used to design well-localized pulses for different lattice structures \cite{blt}.
Recently, the use of Hermite functions has seen renewed interest in various OTFS applications. For example, the time-frequency localized pulse designed and solved in \cite{mc_haas_tfl} as a linear combination of Hermite functions has been used to mitigate fractional delay effects in discrete Zak transform based OTFS (DZT-OTFS) in \cite{dzt_tfl}. Also, individual single-order Hermite functions have been employed 
for multiplexing of users in OTFS-NOMA in \cite{mc_otfs_tfl}. We select the Hermite functions in our work since they provide a flexible design framework. Hermite functions are a set of mutually orthogonal functions, each formed by the product of a Hermite polynomial and a Gaussian envelope. The set of Hermite functions forms a complete orthonormal basis for $L^2(\mathbb{R})$, which is the key to the flexibility of the framework, as it allows the construction of a desired pulse shape by choosing the appropriate coefficients for the linear combination. Furthermore, these basis functions are considered the most time-frequency localized orthonormal basis for finite-energy signals \cite{blt}. This optimal localization stems from their foundation on the Gaussian pulse, the unique function that achieves the minimum of the Heisenberg uncertainty principle. This unique duality—a complete orthonormal basis composed of individually localized functions—makes the Hermite set a good choice for systematically engineering high-performance DD pulses. The new contributions in this paper are summarized as follows:
\begin{itemize}
\item We propose a systematic and novel pulse design framework for Zak-OTFS based on a linear combination of Hermite basis functions that offers greater design flexibility in terms of control of ISI and sidelobe energies. The proposed pulse design does not incur bandwidth or time expansion.
\item We formulate the coefficient design as a constrained optimization problem with ISI energy as the objective to minimize, which we solve efficiently using singular value decomposition (SVD).
\item We quantitatively evaluate the designed pulse's ISI and sidelobe energies as a function of the number of basis functions used in the construction, providing key insights into the trade-off between localization and orthogonality. 
\item For the proposed class of Hermite pulses, we derive closed-form expressions for the I/O relation and noise covariance in Zak-OTFS. The derivation transforms computationally intensive integrals into finite summations by leveraging the ambiguity function and Fourier eigenfunction properties of the Hermite basis, which significantly reduces simulation run times.
\item Through extensive simulations of Zak-OTFS with embedded pilot and model-free I/O relation estimation in Vehicular-A channels \cite{ITU_VehA} with fractional DDs, we demonstrate that the proposed optimized pulse significantly outperforms the canonical sinc and Gaussian pulses in terms of bit error rate (BER) and achieves a BER performance which is on par with that of  the state-of-the-art GS pulse, validating the proposed systematic and flexible DD pulse design framework for Zak-OTFS. 
\end{itemize}
The rest of the paper is organized as follows. Section \ref{sec:system_model} introduces the Zak-OTFS system model. Section \ref{sec:proposed_filter} presents the proposed pulse design including the formulation and solution of the constrained optimization problem, and the derivation of closed-form expressions for I/O relation and noise covariance. Simulation results and discussions are presented in Sec. \ref{sec:results}.  Conclusions and future 
work are presented in Sec. \ref{sec:concl}.

\section{Zak-OTFS system model}
\label{sec:system_model}
\begin{figure*}
\centering    \includegraphics[width=0.95\linewidth]{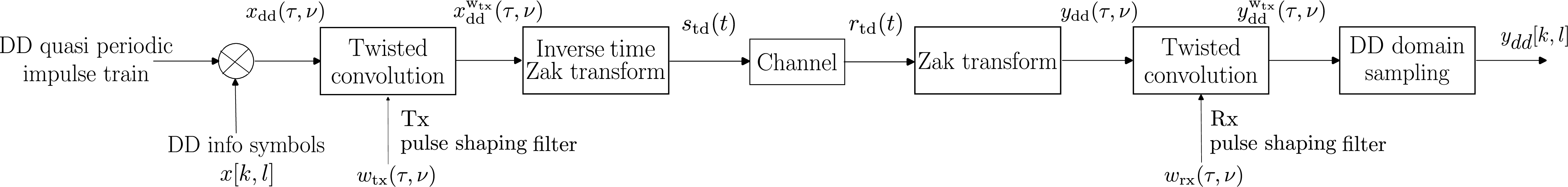}
\vspace{-1mm}
\caption{Block diagram of Zak-OTFS transceiver.}
\label{fig1}      
\vspace{-4mm}
\end{figure*}

Figure \ref{fig1} shows the block diagram of a Zak-OTFS transceiver.
In Zak-OTFS, a pulse in the DD domain is the basic information carrier. A DD pulse is a quasi-periodic localized function defined by a delay period $\tau_{\mathrm{p}}$ and a Doppler period $\nu_{\mathrm{p}}=\frac{1}{\tau_{\mathrm{p}}}$. The fundamental period in the DD domain is defined as 
$\mathcal{D}_{0}= \{(\tau,\nu): 0\leq\tau<\tau_{\mathrm p}, 0\leq\nu<\nu_{\mathrm p}\}$,
where $\tau$ and $\nu$ represent the delay and Doppler variables, respectively. The fundamental period is discretized into $M$ bins on the delay axis and $N$ bins on the Doppler axis, as 
$\big\{(k\frac{\tau_{{\mathrm p}}}{M},l\frac{\nu_{{\mathrm p}}}{N}) | k=0,\ldots,M-1,l=0,\ldots,N-1\big\}$. The time domain Zak-OTFS frame is limited to a time duration $T=N\tau_{\mathrm p}$ and a bandwidth $B=M\nu_{\mathrm p}$. In each frame, $MN$ information symbols drawn from a modulation alphabet ${\mathbb A}$, $x[k,l]\in {\mathbb A}$, $k=0,\ldots,M-1$, $l=0,\ldots,N-1$, are multiplexed in the DD domain. The $x[k,l]$s are encoded according to the following equation to obtain a quasi-periodic extension of the signal in the discrete DD domain:
\begin{equation}
x_{\mathrm {dd}}[k+nM,l+mN] = x[k,l]e^{j2\pi n \frac{l}{N}}, \ n,m\in\mathbb{Z}.
\label{quasi_per}
\end{equation}
These discrete DD domain signals $x_{\mathrm{dd}}[k,l]$s are supported on the information lattice 
$\Lambda_{\mathrm{dd}}=
\big\{\big(k\frac{\tau_{\mathrm p}}{M},l\frac{\nu_{\mathrm p}}{N}\big) | k,l\in \mathbb{Z}\big\}$. The continuous DD domain information signal is given by
\begin{equation}
x_{\mathrm{dd}}(\tau,\nu)=\sum_{k,l\in \mathbb{Z}} x_{\mathrm{dd}}[k,l] \delta\Big(\tau-\frac{k\tau_{\mathrm p}}{M}\Big)\delta\Big(\nu-\frac{l\nu_{\mathrm p}}{N}\Big),
\end{equation}
where $\delta(.)$ denotes the Dirac-delta impulse function. For any $n,m\in \mathbb{Z}$, we have
$x_{\mathrm{dd}}(\tau+n\tau_{\mathrm{p}},\nu+m\nu_{\mathrm{p}})=e^{j2\pi n\nu \tau_{\mathrm{p}}}x_{\mathrm{dd}}(\tau,\nu)$,
so that $x_{\mathrm{dd}}(\tau,\nu)$ is periodic with period $\nu_{\mathrm p}$ along the Doppler axis and quasi-periodic with period $\tau_{\mathrm p}$ along the delay axis.

The DD domain transmit signal $x_{\mathrm{dd}}^{w_{\mathrm{tx}}}(\tau,\nu)$ is given by the twisted convolution of the transmit pulse shaping filter $w_{\mathrm{tx}}(\tau,\nu)$ with $x_{\mathrm{dd}}(\tau,\nu)$ as $x_{\mathrm{dd}}^{w_{\mathrm{tx}}}(\tau,\nu) = w_{\mathrm{tx}}(\tau,\nu)*_{\sigma}x_{\mathrm{dd}}(\tau,\nu)$, where $*_{\sigma}$ denotes the twisted convolution\footnote{Twisted convolution of two DD functions $a(\tau,\nu)$ and $b(\tau,\nu)$ is defined as $a(\tau,\nu) \ast_\sigma b(\tau,\nu) \overset{\Delta}{=} \iint a(\tau', \nu') b(\tau-\tau',\nu-\nu')e^{j2\pi\nu'(\tau-\tau')}d\tau'  d\nu'$.}. The transmitted time domain (TD) signal $s_{\mathrm{td}}(t)$ is the TD realization of $x_{\mathrm{dd}}^{w_{\mathrm{tx}}}(\tau,\nu)$, given by
$s_{\mathrm{td}}(t)=Z_{t}^{-1}\left(x_{\mathrm{dd}}^{w_{\mathrm{tx}}}(\tau,\nu)\right)$, where $Z_{t}^{-1}$ denotes the inverse time-Zak transform operation\footnote{Inverse time-Zak transform of a DD function $a(\tau,\nu)$ is defined as $Z_{t}^{-1}(a(\tau,\nu)) \overset{\Delta}{=} \sqrt{\tau_{\mathrm p}} \int_0^{\nu_{\mathrm p}} a(t,\nu) d\nu$.}. The transmit pulse shaping filter $w_{\mathrm{tx}}(\tau,\nu)$ limits the time and bandwidth of the transmitted signal $s_{\mathrm{td}}(t)$. The transmit signal $s_{\mathrm{td}}(t)$ passes through a doubly-selective channel, resulting in the output signal $r_{\mathrm{td}}(t)$. The DD domain impulse response of the physical channel $h_{\mathrm{phy}}(\tau,\nu)$ is given by
\begin{equation}
h_{\mathrm{phy}}(\tau,\nu)=\sum_{i=1}^{P}h_{i}\delta(\tau-\tau_{i})\delta(\nu-\nu_{i}),
\end{equation}
where $P$ denotes the number of DD paths, and the $i$th path has gain $h_{i}$, delay shift $\tau_{i}$, and Doppler shift $\nu_{i}$. 

The received TD signal $y(t)$ at the receiver is given by $y(t)=r_{\mathrm{td}}(t)+n(t)$, where $n(t)$ is AWGN with variance $N_{0}$, i.e., $\mathbb{E}[n(t)n(t+t')]=N_{0}\delta(t')$. The TD signal $y(t)$ is converted to the corresponding DD domain signal $y_{\mathrm{dd}}(\tau,\nu)$ by applying Zak transform\footnote{Zak transform of a continuous TD signal $a(t)$ is defined as
$Z_t\left(a(t)\right) \overset{\Delta}{=} \sqrt{\tau_p} \sum_{k \in \mathbb{Z}} a(\tau + k \tau_{\mathrm p}) e^{-j2\pi\nu k\tau_{\mathrm p}}$.}, i.e.,
\begin{eqnarray}
y_{\mathrm{dd}}(\tau,\nu) = Z_{t}(y(t)) 
= r_{\mathrm{dd}}(\tau,\nu)+n_{\mathrm{dd}}(\tau,\nu),
\end{eqnarray}
where $r_{\mathrm{dd}}(\tau,\nu)=h_{\mathrm{phy}}(\tau,\nu)*_{\sigma}w_{\mathrm{tx}}(\tau,\nu)*_{\sigma}x_{\mathrm{dd}}(\tau,\nu)$ is the Zak transform of $r_{\mathrm{td}}(t)$, given by the twisted convolution cascade of $x_{\mathrm{dd}}(\tau,\nu)$, $w_{\mathrm{tx}}(\tau,\nu)$, and $h_{\mathrm{phy}}(\tau,\nu)$,  and $n_{\mathrm{dd}}(\tau,\nu)$ is the Zak transform of $n(t)$. The receive filter is matched to the transmit pulse, i.e., $w_{\mathrm{rx}}(\tau,\nu) = w_{\mathrm{tx}}^*(-\tau,-\nu)e^{j2\pi\nu\tau}$ \cite{zak_otfs6}-\cite{zak_otfs11} and acts on $y_{\mathrm{dd}}(\tau,\nu)$ through twisted convolution to give the output 
\begin{eqnarray}
\hspace{-4mm} 
y_{\mathrm{dd}}^{w_{\mathrm{rx}}}(\tau,\nu) & \hspace{-2mm} = & \hspace{-2mm} w_{\mathrm{rx}}(\tau,\nu)*_{\sigma}y_{\mathrm{dd}}(\tau,\nu) \nonumber \\ 
& \hspace{-15mm} = & \hspace{-8mm} \underbrace{w_{\mathrm{rx}}(\tau,\nu)*_{\sigma}h_{\mathrm{phy}}(\tau,\nu)*_{\sigma}w_{\mathrm{tx}}(\tau,\nu)}_{\overset{\Delta}{=} \ h_{\mathrm{eff}}(\tau,\nu)}*_{\sigma}x_{\mathrm{dd}}(\tau,\nu) \nonumber \\ 
&\hspace{-15mm} & \hspace{-8mm} + \ \underbrace{w_{\mathrm{rx}}(\tau,\nu)*_{\sigma}n_{\mathrm{dd}}(\tau,\nu)}_{\overset{\Delta}{=} \ n_{\mathrm{dd}}^{w_{\mathrm{rx}}}(\tau,\nu)}, 
\label{cont1}
\end{eqnarray}
where $h_{\mathrm{eff}}(\tau,\nu)$ denotes the `effective channel' consisting of the twisted convolution cascade of $w_{\mathrm{tx}}(\tau,\nu),\ h_{\mathrm{phy}}(\tau,\nu)$, and $w_{\mathrm{rx}}(\tau,\nu)$, and $n_{\mathrm{dd}}^{w_{\mathrm{rx}}}(\tau,\nu)$ denotes the noise filtered through the receive filter. The DD signal $y_{\mathrm{dd}}^{w_{\mathrm{rx}}}(\tau,\nu)$ is sampled on the information grid, resulting in the discrete quasi-periodic DD domain received signal $y_{\mathrm{dd}}[k,l]$ as
\begin{eqnarray}
\hspace{-4mm}y_{\mathrm{dd}}[k,l]&\hspace{-2mm}=&\hspace{-2mm} y_{\mathrm{dd}}^{w_{\mathrm{rx}}}\left(\tau=\frac{k\tau_{\mathrm p}}{M},\nu=\frac{l\nu_{\mathrm p}}{N}\right) \nonumber \\
&\hspace{-26mm}=&\hspace{-17
mm} \sum_{k',l'\in\mathbb{Z}} \hspace{-2mm} h_{\mathrm{eff}}[k-k',l-l']x_{\mathrm{dd}}[k',l'] e^{j2\pi\frac{k'(l-l')}{MN}} + n_{\mathrm{dd}}[k,l].
\label{eq_new2}
\end{eqnarray}
Hence, the $y_{\mathrm{dd}}[k,l]$ samples are given by $y_{\mathrm{dd}}[k,l]=h_{\mathrm{eff}}[k,l]*_{\sigma\text{d}}x_{\mathrm{dd}}[k,l]+n_{\mathrm{dd}}[k,l]$,
where $*_{\sigma\text{d}}$ is twisted convolution in discrete DD domain, i.e., 
$h_{\mathrm{eff}}[k,l]*_{\sigma\text{d}}x_{\mathrm{dd}}[k,l] = \sum_{k',l'\in\mathbb{Z}}h_{\mathrm{eff}}[k-k',l-l']x_{\mathrm{dd}}[k',l'] e^{j2\pi\frac{k'(l-l')}{MN}}$, where the effective channel filter $h_{\mathrm{eff}}[k,l]$ and filtered noise samples $n_{\mathrm{dd}}[k,l]$ are given by
\begin{align}
h_{\text{eff}}[k,l]=h_{\text{eff}}\left(\tau=\frac{k\tau_{p}}{M},\nu=\frac{l\nu_{p}}{N}\right), \label{discr2} \\ 
n_{\text{dd}}[k,l]=n_{\text{dd}}^{w_{\mathrm{rx}}}\left(\tau=\frac{k\tau_{p}}{M},\nu=\frac{l\nu_{p}}{N}\right).
\label{discr3}
\end{align}
Because of the quasi-periodicity in the DD domain, it is sufficient to consider the received samples $y_{\mathrm{dd}}[k,l]$ within the fundamental period $\mathcal{D}_0$. Writing the $y_{\mathrm{dd}}[k,l]$ samples as a vector, the received signal model can be written in matrix-vector form as \cite{zak_otfs1},\cite{zak_otfs2}
\begin{equation}
\mathbf{y}=\mathbf{H_\text{eff}x}+\mathbf{n},
\label{sys_mod}
\end{equation}
where $\mathbf{x,y,n} \in\mathbb{C}^{MN\times 1}$, such that their $(kN+l+1)$th entries are given by $x_{kN+l+1}=x_{\mathrm{dd}}[k,l]$, $y_{kN+l+1}=y_{\mathrm{dd}}[k,l]$, $n_{kN+l+1}=n_{\mathrm{dd}}[k,l]$, and $\mathbf{H}_{\text{eff}}\in\mathbb{C}^{MN\times MN}$ is the effective channel matrix such that
\vspace{-1mm}
\begin{eqnarray}
\mathbf{H}_\text{eff}[k'N+l'+1,kN+l+1] & \hspace{-2mm} = & \hspace{-2mm} \sum_{m,n\in\mathbb{Z}}h_{\mathrm{eff}}[k'-k-nM, \nonumber \\
& \hspace{-45mm} & \hspace{-35mm} l'-l-mN]e^{j2\pi nl/N}e^{j2\pi\frac{(l'-l-mN)(k+nM)}{MN}},
\label{eqn_channel_matrix}
\vspace{-4mm}
\end{eqnarray}
where $k',k=0,\ldots,M-1$, $l',l=0,\ldots,N-1$. At the receiver, the I/O relation captured in (\ref{sys_mod}) needs to be estimated for data detection. I/O relation estimation refers to the task of estimating the $h_{\mathrm{eff}}[k,l]$ coefficients at the receiver and constructing an estimate of the ${\bf H}_{\mathrm{eff}}$ matrix using (\ref{eqn_channel_matrix}), which is used for data detection. 

\subsection{DD pulse shaping filters} 
In the absence of pulse shaping, i.e., $w_\text{tx}(\tau,\nu)=\delta(\tau,\nu)$, the transmit signal has infinite time duration and bandwidth. Pulse shaping limits the time and bandwidth of transmission. We consider transmit DD pulse shaping filters of the form $w_\text{tx}(\tau,\nu)=w_1(\tau)w_2(\nu)$ \cite{zak_otfs2},\cite{zak_otfs6}. The time duration $T'$ of each frame is approximately related to the spread of $w_2({\nu})$ along the Doppler axis as $\frac{1}{T'}$. Similarly, the bandwidth $B'$ is approximately related to the spread of $w_1(\tau)$ along the delay axis as $\frac{1}{B'}$. That is, a larger bandwidth and time duration implies a smaller DD spread of $w_\text{tx}(\tau,\nu)$, and hence a smaller contribution to the spread of $h_\text{eff}(\tau,\nu)$. Sinc, root-raised cosine (RRC), Gaussian, and Gaussian-sinc (GS) filters have been considered in the Zak-OTFS literature.
RRC filter incurs expansion in bandwidth/time depending on the roll-off factor. Sinc, Gaussian, and GS filters do no incur bandwidth or time expansion. We consider filters which do not incur bandwidth or time expansion. The sinc, Gaussian, and GS filters are described below. 

{\em Sinc filter:}
The DD domain sinc filter is given by
\begin{equation}
w_\text{tx}(\tau,\nu)
=\underbrace{\sqrt{B}\text{sinc}(B\tau)}_{w_1(\tau)} \underbrace{\sqrt{T}\text{sinc}(T\nu)}_{w_2(\nu)}. 
\label{eq:sinc1}
\end{equation}
For sinc filter, the frame duration $T'=T$ and frame bandwidth $B'=B$ (i.e., there is no time or bandwidth expansion), resulting in a spectral efficiency of $\frac{BT}{B'T'}=1$ symbol/dimension. Sinc filter has the advantage of nulls at the sampling points, which is optimum for data detection with perfect channel knowledge. However, it has high sidelobes which is detrimental for I/O relation estimation.

{\em Gaussian filter:}
The DD domain Gaussian filter is given by 
\begin{equation}
\hspace{-2mm}
w_\text{tx}(\tau,\nu) \hspace{-0.5mm} = \hspace{-0.5mm} \underbrace{\left(\frac{2\alpha_{\tau}B^2}{\pi}\right)^{\frac{1}{4}}e^{-\alpha_{\tau}B^{2}\tau^{2}}}_{w_1(\tau)} \ \underbrace{\left(\frac{2\alpha_{\nu}T^2}{\pi}\right)^{\frac{1}{4}}e^{-\alpha_{\nu}T^{2}\nu^{2}}}_{w_2(\nu)}\hspace{-1mm}, \hspace{-0mm}
\label{eq:gauss1}
\end{equation}
which can be configured by adjusting the parameters $\alpha_{\tau}$ and $\alpha_{\nu}$.
Due to the infinite support in Gaussian pulse, its effective time duration $T'$ and bandwidth $B'$ are defined by 99$\%$ energy containment. To achieve no time and bandwidth expansion (i.e., $T'=T$, $B'=B$), the roll-off factors are set to $\alpha_{\tau}=\alpha_{\nu}=1.584$. Gaussian filter has very low sidelobes, because of which it is attractive for I/O relation estimation. However, its lack of nulls at sampling points cause ISI leading to poor detection performance. 

{\em GS filter:}
For GS filter, $w_1({\tau})$ and $w_2({\nu})$ are given by
$w_{1}(\tau)=\Omega_{\tau}\sqrt{B}\mathrm{sinc}(B\tau)e^{-\alpha_{\tau}B^{2}\tau^{2}}$ and $w_{2}(\nu)=\Omega_{\nu}\sqrt{T}\mathrm{sinc}(T\nu)e^{-\alpha_{\nu}T^{2}\nu^{2}}$,
so that 
\begin{equation}
w_\text{tx}(\tau,\nu)=\Omega_\tau\Omega_\nu\sqrt{BT} \mathrm{sinc}(B\tau)\mathrm{sinc}(T\nu)e^{-\alpha_{\tau}B^{2}\tau^{2}} \hspace{-1mm} e^{-\alpha_{\nu}T^{2}\nu^{2}}\hspace{-1mm}.
\label{gsf}
\end{equation}
The roll-off parameters are chosen to be $\alpha_{\tau}=\alpha_{\nu}=0.044$ for no bandwidth/time expansion and the energy normalization parameters are $\Omega_{\tau}=\Omega_{\nu}=1.0278$ \cite{gs}. The GS filter inherits the complementary strengths of both sinc and Gaussian filters, achieving a better BER performance compared to sinc and Gaussian filters under model-free I/O relation estimation \cite{gs}.

\subsection{Model-free I/O relation estimation}
Zak-OTFS has the advantage of a simple model-free I/O relation estimation in fractional DD channels, when the system is operated in the crystalline region \cite{zak_otfs2}. A system is said to be operating in the crystalline region if the crystallization condition is met. The crystallization condition is said to be met if the delay spread of the effective channel is less than the delay period $\tau_{\mathrm{p}}$ and the
Doppler spread of the effective channel is less than the Doppler period $\nu_{\mathrm{p}}$. Model-free estimation does not require explicit estimation of the channel parameters \{$h_i, \tau_i, \nu_i$\}, $i=1,\ldots,P$, and a simple read-off operation from a suitably defined read-off region around the pilot location in the received frame gives the I/O relation estimate \cite{zak_otfs2},\cite{gs}. 

\section{Proposed Hermite Pulse Shaping Filter}
\label{sec:proposed_filter}
In this section, we present the proposed DD pulse design framework using Hermite basis functions. We begin by introducing the Hermite functions and their excellent localization properties. We develop the structure of the proposed pulse as a scaled linear combination of these basis functions. The core of the framework,  i.e., formulation of a constrained optimization problem (solved via SVD) to derive the optimal coefficients that minimize the ISI energy, is then detailed. Next, we analyze the key characteristics of the optimized pulse, focusing on the trade-off between localization and orthogonality. Finally, for the proposed Hermite pulses, we present the derivation for the closed-form expressions of the I/O relation and noise covariance in Zak-OTFS.

\subsection{Hermite functions as an orthonormal basis}
The challenge in pulse design, as highlighted by the Balian-Low Theorem, is to find a function that simultaneously offers good TF localization and orthogonality. The Hermite functions are particularly well-suited for this challenge. The set of Hermite functions, $\{\psi_n(t)\}_{n=0}^{\infty}$, forms a complete orthonormal basis for the Hilbert space of square-integrable functions $L^2(\mathbb{R})$. This means any practical pulse shape can be accurately represented as a linear combination of Hermite functions. Furthermore, each individual function is highly localized in time and frequency. The Hermite functions are defined using the Hermite polynomials, $H_n(t)$, which are given by the Rodrigues' formula:
\begin{equation}
H_n(t) = (-1)^n e^{t^2} \frac{d^n}{dt^n} e^{-t^2}.
\label{eq:hermite_poly}
\end{equation}
The $n$th order orthonormal Hermite basis function, $\psi_n(t)$, is constructed by multiplying the Hermite polynomial with a Gaussian envelope as
\begin{equation}
\psi_n(t) = \frac{\pi^{-1/4}}{\sqrt{2^n n!}} H_n(t) e^{-t^2/2}.
\label{eq:hermite_basis}
\end{equation}
Each basis function $\psi_n(t)$ is well localized, with the zeroth-order function $\psi_0(t)$ being the Gaussian pulse that achieves the minimum possible TF spread allowed by the Heisenberg uncertainty principle.

\subsection{Construction of the proposed pulse shape}
For practical implementation in a communication system with a given bandwidth and time duration, we introduce a scaled version of the Hermite basis. A real-valued scaling parameter $\sigma > 0$ is used to control the pulse width, which is analogous to a roll-off factor. The scaled orthonormal Hermite basis functions are defined as
\begin{equation}
\phi_n(\sigma, t) = \sqrt{\sigma} \psi_n(\sigma t).
\label{eq:scaled_hermite_basis}
\end{equation}
This transformation scales the function's argument by $\sigma$ and its amplitude by $\sqrt{\sigma}$, a structure that conveniently preserves the orthonormality of the basis set.

We propose to construct the pulse shaping filters for the delay and Doppler axes, $w_1(\tau)$ and $w_2(\nu)$, respectively, as a linear combination of a finite number of these scaled basis functions. To ensure that the resulting pulse is symmetric (an even function), we use only  even-ordered Hermite functions. The proposed pulses are thus defined as
\begin{eqnarray}
w_1(\tau) & \hspace{-2mm} = & \hspace{-2mm} \sum_{n=0}^{N_c-1} c_{2n} \phi_{2n}(\sigma_\tau, \tau), \label{eq:w1_pulse} \\
w_2(\nu) & \hspace{-2mm} = & \hspace{-2mm} \sum_{n=0}^{N_c-1} d_{2n} \phi_{2n}(\sigma_\nu, \nu), \label{eq:w2_pulse}
\end{eqnarray}
where $N_c$ is the number of even-ordered basis functions.
$\{c_{2n}\}_{n=0}^{N_c-1}$ and $\{d_{2n}\}_{n=0}^{N_c-1}$ are the sets of real-valued coefficients that need to be optimized, and are represented by the vectors $\mathbf{c} = [c_0 \ c_2 \ \dots \ c_{2(N_c-1)}]^T \in \mathbb{R}^{N_c \times 1}$ and $\mathbf{d} = [d_0 \ d_2 \ \dots \ d_{2(N_c-1)}]^T \in \mathbb{R}^{N_c \times 1}$, respectively. To ensure energy normalization in the resulting pulses, we impose a unit-norm constraint on these coefficient vectors, i.e., $\|\mathbf{c}\|=\|\mathbf{d}\|=1$. This follows from Parseval's theorem for orthonormal bases, where the pulse energy equals the squared norm of its coefficients \cite{mallat_tb}. This constraint also makes the subsequent optimization problem well-posed. The scaling parameters $\sigma_\tau$ and $\sigma_\nu$ are directly linked to the system parameters and roll-off factors $\beta_\tau$ and $\beta_\nu$ as
\begin{equation}
\sigma_\tau = B\sqrt{2\beta_\tau} \quad \text{and} \quad \sigma_\nu = T\sqrt{2\beta_\nu}.
\label{eq:sigma_def}
\end{equation}

The roll-off factors are critical design parameters that control the pulse's compactness. However, their selection is not independent of the coefficient optimization. The final shape of the pulse, and therefore its energy containment, depends on the coefficients $\{c_{2n}\}$ and $\{d_{2n}\}$, which are themselves derived for a given $\sigma$ (and thus a given $\beta$). To address this interdependency, the values for $\beta_\tau$ and $\beta_\nu$ are determined numerically to find the largest roll-off that satisfies a 99\% in-band energy requirement for the final optimized pulse. This ensures that the pulse parameters are not chosen in isolation but are jointly determined to meet both ISI minimization and spectral containment criteria. We find the minimum $\beta$ that satisfies this via the following procedure: for a candidate value of $\beta$, the optimal coefficient is first computed by solving the ISI minimization problem described in Sec. \ref{subsec:optimal}. The resulting pulse is constructed and its energy containment is checked. This process is repeated in a methodical search to find the maximum $\beta$ that meets the 99\% threshold. This systematic approach guarantees that the resulting pulses are both spectrally efficient and optimized for minimum ISI.

\subsection{Derivation of optimal coefficients}
\label{subsec:optimal}
We formulate the coefficient design problem as a constrained optimization problem with the ISI energy as the objective to minimize. The ISI energy is the energy of the pulse at the non-zero Nyquist sampling instants. 
For the delay domain pulse $w_1(\tau)$ which is even/symmetric by construction, i.e., $w_1(\tau) = w_1(-\tau)$, the ISI energy in $L$ sampling points on either side of the zero sampling point ($\tau=0$) is given by
\begin{eqnarray}
E_{\text{ISI}, w_1} & \hspace{-2mm} = & \hspace{-2mm} 2 \sum_{p=1}^{L} |w_1(p/B)|^2 \nonumber \\
& \hspace{-2mm} = & \hspace{-2mm} 2 \sum_{p=1}^{L} \left| \sum_{n=0}^{N_c-1} c_{2n} \phi_{2n}(\sigma_\tau, p/B) \right|^2 \nonumber \\
& \hspace{-2mm} = & \hspace{-2mm} 2 \|\mathbf{\Phi}_{\tau} \mathbf{c}\|^2.
\label{eq:isi_energy_expanded}
\end{eqnarray}
The last step in (\ref{eq:isi_energy_expanded}) follows from expressing the inner sum as the $p$th element of the vector $\mathbf{\Phi}_{\tau} \mathbf{c}$, where $\mathbf{\Phi}_{\tau} \in \mathbb{R}^{L \times N_c}$ is a matrix with entries $(\Phi_{\tau})_{p,n} = \phi_{2n}(\sigma_\tau, p/B)$, for $p = 1, \dots, L$ and $n = 0, \dots, N_c-1$. We choose a large value of $L > N_c$, leading to a tall matrix $\mathbf{\Phi}_{\tau}$. The design problem is to find the coefficient vector ${\bf c}$ that minimizes the energy $E_{\text{ISI}, w_1}$. This amounts to minimizing $\|\mathbf{\Phi}_{\tau} \mathbf{c}\|^2$ subject to the unit-norm constraint, which is formally stated as
\begin{equation}
\underset{\|\mathbf{c}\|=1}{\text{minimize}} \ \|\mathbf{\Phi}_{\tau} \mathbf{c}\|^2.
\label{eq:optimization_problem}
\end{equation} 
The solution to this constrained optimization problem is given by the following theorem.

\begin{theorem}
Given the SVD of the real matrix $\mathbf{\Phi}_{\tau} = \mathbf{U}_{\tau}\mathbf{\Sigma}_{\tau}\mathbf{V}_{\tau}^T$, 
where $\mathbf{U}_{\tau} \in \mathbb{R}^{L \times L}$ and $\mathbf{V}_{\tau} = [\mathbf{v}_{\tau,1} \ \mathbf{v}_{\tau,2} \ \dots \ \mathbf{v}_{\tau,N_c}] \in \mathbb{R}^{N_c \times N_c}$ are orthogonal matrices, and $\mathbf{\Sigma}_{\tau} \in \mathbb{R}^{L \times N_c}$ is a diagonal matrix whose diagonal entries 
$\sigma_{i,\tau}$ are the singular values of $\mathbf{\Phi}_{\tau}$ arranged in descending order ($\sigma_{1,\tau} \ge \sigma_{2,\tau} \ge \dots \ge \sigma_{N_c,\tau} \ge 0$).
Then the optimal coefficient vector $\mathbf{c}_{\mathrm{opt}}$ that solves the constrained ISI energy minimization problem in (\ref{eq:optimization_problem}) is the last right singular vector of the matrix $\mathbf{\Phi}_{\tau}$, i.e.,
\begin{equation}
\mathbf{c}_{\mathrm{opt}} = \mathbf{v}_{\tau,N_c}.
\end{equation}
Furthermore, the minimum ISI energy achieved by the resulting pulse is
$E_{\text{ISI}, w_1}^{\min} = 2\sigma_{N_c,\tau}^2$.
\label{thm:svd}
\end{theorem}

\begin{IEEEproof}
See Appendix \ref{app:svd}.
\end{IEEEproof}

In a similar way, for the Doppler domain pulse $w_2(\nu)$, the design objective is to minimize the ISI energy, which is concentrated at the Nyquist sampling instants $q/T$ for integers $q \neq 0$. The ISI energy expression is analogous to (\ref{eq:isi_energy_expanded}) and is given by
\begin{eqnarray}
E_{\text{ISI}, w_2} & \hspace{-2mm} = & \hspace{-2mm} 2 \sum_{q=1}^{L} |w_2(q/T)|^2 \nonumber \\
& \hspace{-2mm} = & \hspace{-2mm} 2 \sum_{q=1}^{L} \left| \sum_{n=0}^{N_c-1} d_{2n} \phi_{2n}(\sigma_\nu, q/T) \right|^2 \nonumber \\
& \hspace{-2mm} = & \hspace{-2mm} 2 \|\mathbf{\Phi}_{\nu} \mathbf{d}\|^2.
\label{eq:isi_energy_doppler}
\end{eqnarray}
The matrix $\mathbf{\Phi}_{\nu} \in \mathbb{R}^{L \times N_c}$ is constructed with entries $(\Phi_{\nu})_{q,n} = \phi_{2n}(\sigma_\nu, q/T)$, for $q = 1, \dots, L$ and $n = 0, \dots, N_c-1$.
The optimization problem for the Doppler coefficient vector $\mathbf{d}$ can be stated as
\begin{equation}
\underset{\|\mathbf{d}\|=1}{\text{minimize}} \ \|\mathbf{\Phi}_{\nu} \mathbf{d}\|^2.
\label{eq:optimization_problem_doppler}
\end{equation}
As in the case of obtaining the optimal  vector $\mathbf{c}_{\text{opt}}$, the optimal vector $\mathbf{d}_{\text{opt}}$ is found from the SVD of $\mathbf{\Phi}_{\nu} = \mathbf{U}_{\nu}\mathbf{\Sigma}_{\nu}\mathbf{V}_{\nu}^T$. The optimal coefficient vector is the last right singular vector, $\mathbf{v}_{\nu,N_c}$, corresponding to the smallest singular value $\sigma_{N_c,\nu}$, i.e.,
\begin{equation}
\mathbf{d}_{\text{opt}} = \mathbf{v}_{\nu,N_c},
\label{eq:d_opt_solution}
\end{equation}
and the minimum ISI energy for the optimum Doppler domain pulse is $E_{\text{ISI}, w_2}^{\min} = 2\sigma_{N_c,\nu}^2$. The optimized overall DD domain transmit pulse $w_{\mathrm{tx}}(\tau,\nu)$ is obtained as the product of the optimized $w_1({\tau})$ and $w_2(\nu)$.

\subsection{Analysis of the characteristics of the optimized pulse}
\label{sec:pulse_chara_analysis}
In this section, we analyze the key characteristics of the proposed Hermite pulses designed for different $N_c$ values without incurring bandwidth/time expansion. 
The analysis focuses on the impact of $N_c$ on the fundamental trade-off between orthogonality (quantified by ISI energy) and localization (quantified by sidelobe energy). To illustrate this relationship, we present the heatmaps of the optimized Hermite pulse shapes for different $N_c$, followed by a quantitative evaluation of their energy distribution.

\begin{figure}
\centering \includegraphics[width=1.0\linewidth]{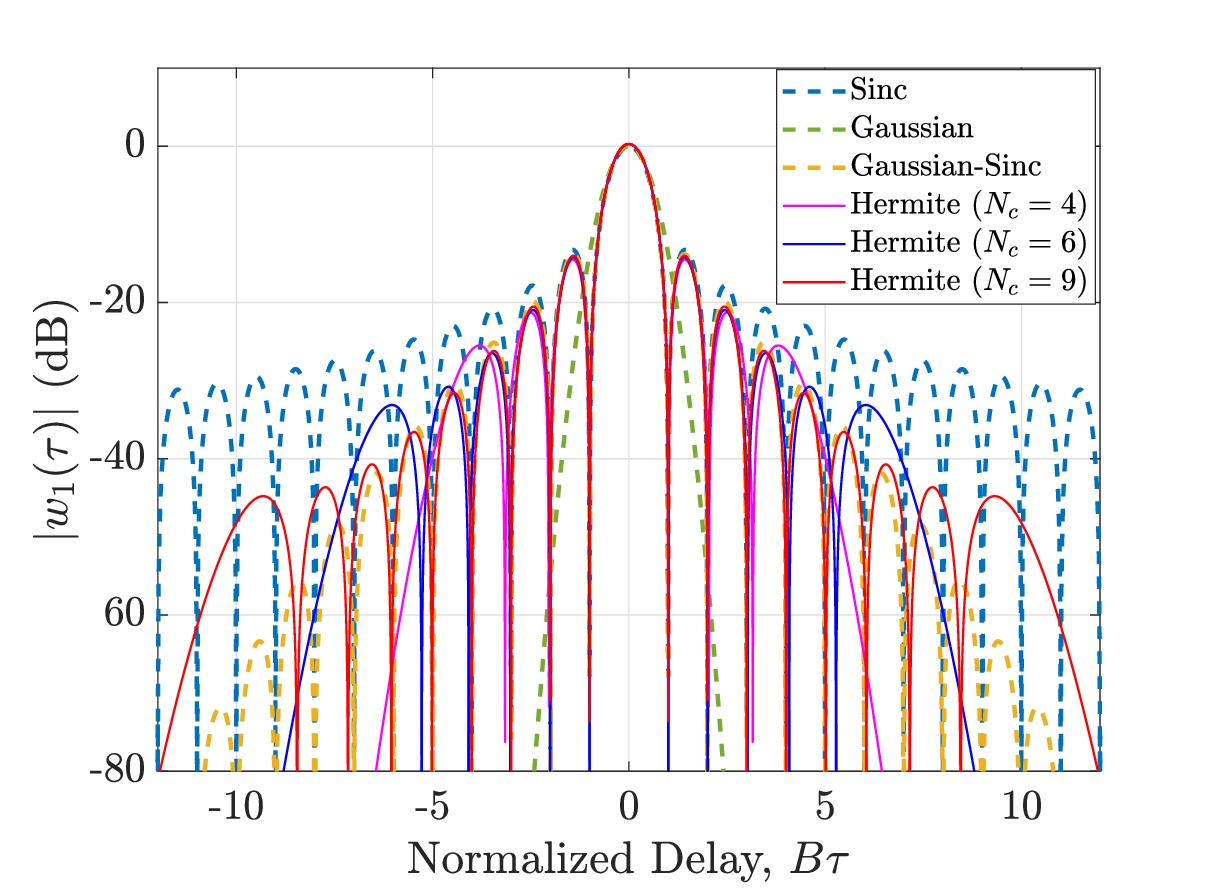}
\caption{Sinc, Gaussian, Gaussian-sinc, and proposed Hermite pulse shapes.}
\label{fig:pulse_shape}
\vspace{-4mm}
\end{figure}

Figure \ref{fig:pulse_shape} shows the delay pulse magnitude $|w_1(\tau)|$ (in dB scale) as a function of the normalized delay $B\tau$. The figure compares the proposed Hermite pulses for different $N_c$ values ($N_c=4,6,9$) with baseline pulses including sinc, Gaussian, and Gaussian-sinc pulses. 
It can be seen that increasing $N_c$ improves orthogonality by extending the number of effective nulls at the Nyquist sampling points. For example, the pulse for $N_c=9$ maintains its critical null structure up to the seventh sampling instant from $\tau=0$, while the energy of the pulse for $N_c=4$ decays rapidly beyond the third sampling instant. Furthermore, the first major sidelobe that appears after this primary region of nulls is significantly higher for smaller $N_c$. This peak is at approximately -25 dB for $N_c=4$, while for $N_c=9$ it is suppressed to below -40 dB. This enhanced orthogonality comes at a direct cost to localization. As $N_c$ increases, the overall energy envelope of the pulse becomes significantly wider, signifying poorer energy containment. This is evident as the pulse for $N_c=9$ is spread across a much larger range of delays compared to the more compact shape of the $N_c=4$ pulse. Figure \ref{fig:pulse_shape} therefore captures the fundamental trade-off: a larger $N_c$ provides superior orthogonality at the expense of localization. 

\begin{figure}
\centering \includegraphics[width=1.0\linewidth]{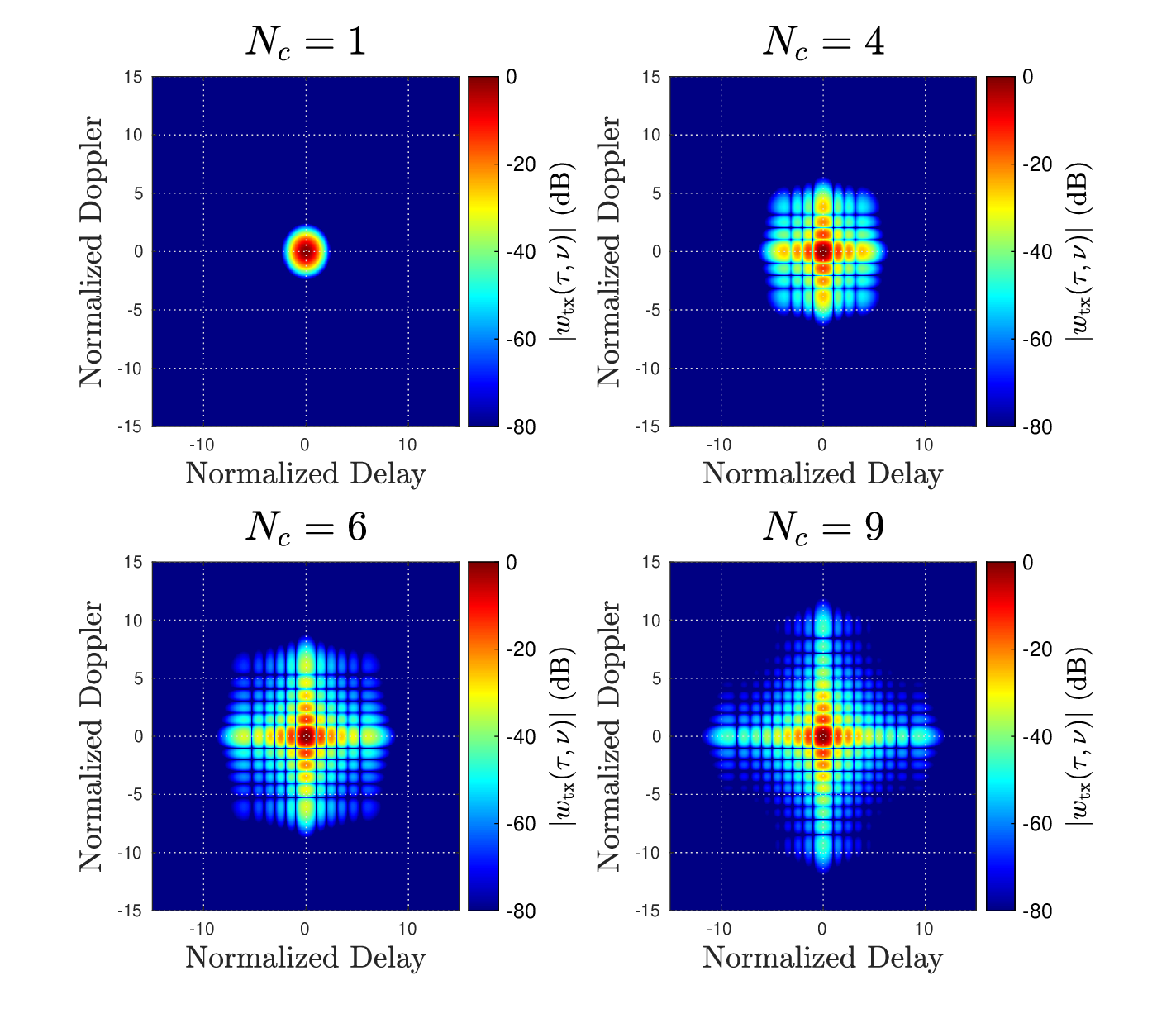}
\caption{Heatmaps of the proposed Hermite pulses for different values of $N_c$. $N_c=1$ corresponds to the canonical Gaussian pulse.}
\label{fig:heatmap}
\vspace{-4mm}
\end{figure}

The above trade-off behavior naturally extends to the DD pulse $w_\text{tx}(\tau,\nu) = w_1(\tau)w_2(\nu)$, as captured in heatmap form in Fig. \ref{fig:heatmap}. This figure illustrates how the optimization process fundamentally redistributes the pulse's energy in the DD plane to promote orthogonality. For the case of $N_c=1$, the pulse is the canonical Gaussian pulse with its energy optimally concentrated around the origin, which has the best localization but lacks the nulls required for orthogonality. As $N_c$ increases, the ISI minimization carves out a sharp grid of nulls along the delay and Doppler axes—the visual signature of a pulse designed for low ISI energy. This process of suppressing energy at the Nyquist sampling points inherently forces that energy to be distributed resulting in the formation of a prominent cross-shaped sidelobe structure. As shown in the progression from $N_c=4$ to $N_c=9$, this structure becomes more pronounced and spreads further from the origin, providing a clear visual illustration that improved orthogonality is achieved by sacrificing the pulse's compact localization.

\begin{figure}
\centering \includegraphics[width=0.975\linewidth]{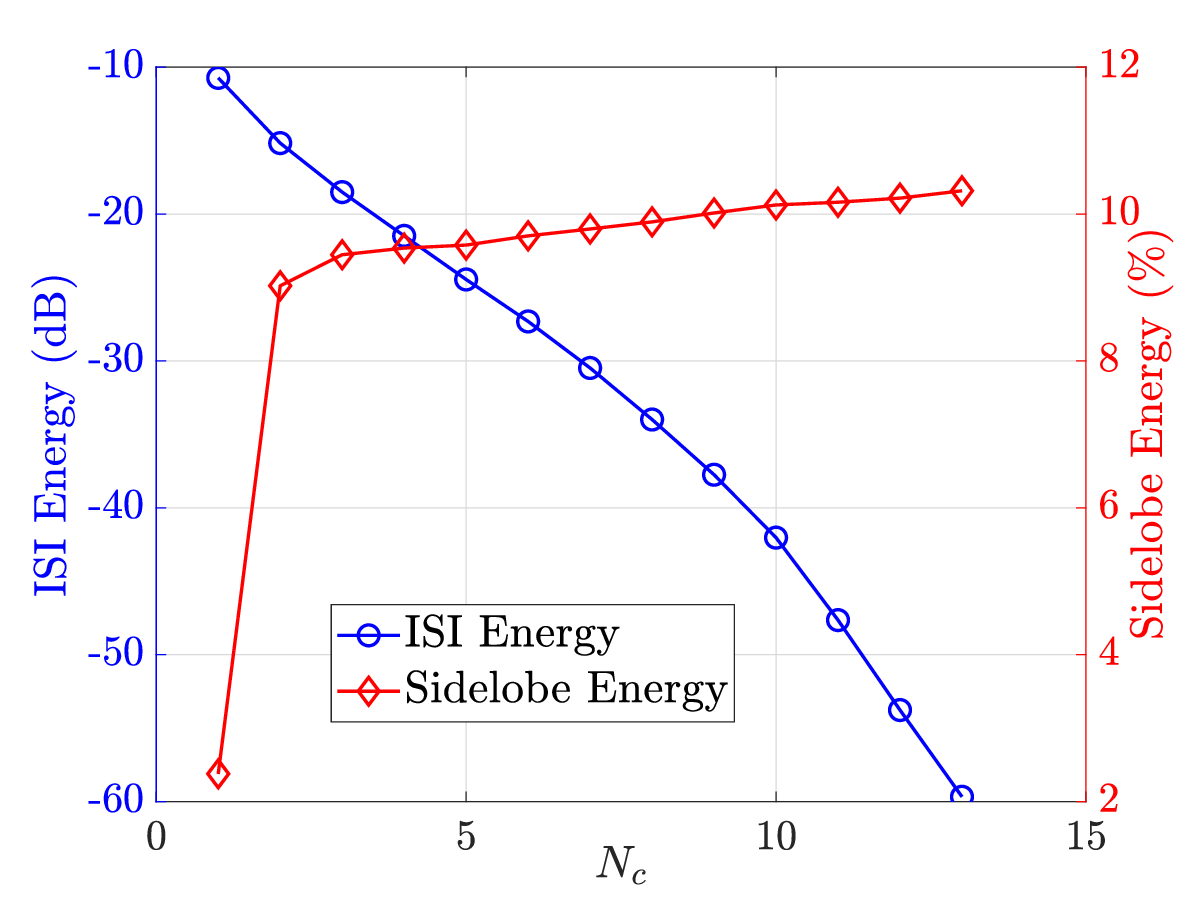}
\caption{ISI energy and sidelobe energy of the proposed Hermite pulses as a function of $N_c$.}
\label{fig:energy_comparison}
\vspace{-4mm}
\end{figure}

In order to quantify the trade-off pictorially observed in Figs. \ref{fig:pulse_shape} and \ref{fig:heatmap},  
we evaluate the two key metrics plotted in Fig. \ref{fig:energy_comparison}: ISI energy (blue curve) to quantify orthogonality and sidelobe energy (red curve) to quantify localization as a function of $N_c$.
The ISI energy is computed analytically for the delay pulse using the result from Theorem \ref{thm:svd}, which states that the minimum achievable ISI energy is directly given by $2\sigma_{N_c}^2$. The effectiveness of this optimization is clear from the blue curve, which shows that the ISI energy drops from approximately -10 dB at $N_c=1$ to -40 dB at $N_c=9$, a large improvement of 30 dB. This confirms that adding more basis functions allows us to create progressively deeper and more effective nulls. The sidelobe energy is numerically computed for the complete DD pulse, $w_\text{tx}(\tau,\nu)$, to capture the total energy leakage in the DD plane, as the energy outside the region $\left[-\frac{\tau_{p}}{M}, \frac{\tau_{p}}{M}\right) \times \left[-\frac{\nu_{p}}{N}, \frac{\nu_{p}}{N}\right)$.  The red curve shows that the gain in orthogonality comes at a direct cost to localization, as the sidelobe energy increases sharply from a low of 2\% at $N_c=1$ to over 10\% at $N_c=9$.

The above analysis demonstrates that the choice of $N_c$ is not arbitrary, simply increasing its value to improve orthogonality is suboptimal due to the detrimental effect on localization. The selection of $N_c$ must, therefore, be based on balancing this trade-off to achieve the best overall system performance. We determine the optimal value based on the BER performance (see Fig. \ref{fig:ber_nmse_vs_Nc} presented in Sec. \ref{sec:results}) and this value is used in subsequent BER simulations.

\subsection{I/O relation and noise covariance closed-form expressions}
\label{sec:analytical_expressions}
To facilitate the theoretical analysis and simulation of Zak-OTFS with the proposed Hermite pulse, we have derived closed-form expressions for the I/O relation and the noise covariance. The derivation transforms computationally intensive integrals into finite summations by leveraging the ambiguity function and Fourier eigenfunction properties of the Hermite basis. These closed-form expressions reduce simulation run times significantly. For the system parameters in Fig. \ref{fig:ber_comparison}, the simulation run times taken for the computation of $h_\text{eff}[k,l]$s with closed-form expression and without closed-form expression (i.e., with numerical integration) for 1000 channel realizations at a given SNR are found to be 14.8 min and 37.4 min, respectively\footnote{The simulations are run on a PC with a 13th Gen Intel Core i7-13700 processor (16 cores, 24 threads) and 48.0 GiB of RAM, using MATLAB R2023b.}. The derived closed-form expressions are presented in the following Theorems \ref{thm:heff} and \ref{thm:noise_cov}, with their corresponding derivations detailed in the Appendices \ref{app:heff} and \ref{app:noise}, respectively.

\begin{theorem}
The DD domain effective channel $h_{\mathrm{eff}}(\tau,\nu)$ for the proposed Hermite pulse is given by 
\begin{eqnarray}
\hspace{-5mm}
h_{\mathrm{eff}}(\tau,\nu) & \hspace{-2mm} = & \hspace{-2mm} e^{j\pi\nu\tau} \sum_{i=1}^{P} h_i e^{-j\pi\nu_i\tau_i} \nonumber \\
& \hspace{-35mm} & \hspace{-20mm} \times \left( \sum_{n_1,m_1=0}^{N_c-1} \hspace{-2mm} c_{2n_1}c_{2m_1} A_{\psi_{2n_1},\psi_{2m_1}}\left(\sigma_\tau(\tau_i-\tau), -\frac{\nu_i}{\sigma_\tau}\right) \right) \nonumber \\
& \hspace{-35mm} & \hspace{-20mm} \times \left( \sum_{n_2,m_2=0}^{N_c-1} \hspace{-2mm} d_{2n_2}d_{2m_2} A_{\psi_{2n_2},\psi_{2m_2}}\left(\sigma_\nu(\nu_i-\nu), \frac{\tau}{\sigma_\nu}\right) \right),
\label{eq:heff_closed_form}
\end{eqnarray}
where $A_{\psi_n,\psi_m}(\cdot, \cdot)$ is the symmetric cross-ambiguity function of the Hermite basis functions, given by \cite{herm_amb}
\begin{eqnarray}
A_{\psi_n,\psi_m}(\tau, \nu) & \hspace{-2mm} = & \hspace{-2mm} \pi^{\frac{1}{4}} \sqrt{\frac{n!m!}{2^{n+m}}} (-j)^{n+m} e^{j(n-m) \tan^{-1}\left(\frac{\tau}{2\pi\nu}\right)} \nonumber \\ 
& \hspace{-35mm} & \hspace{-20mm} \times \sum_{k=0}^{\lfloor\frac{m+n}{2}\rfloor} v_{k,n,m} \psi_{n+m-2k}\left(\sqrt{\frac{\tau^2+(2\pi\nu)^2}{2}}\right), 
\label{eq:ambiguity_closed_form}
\end{eqnarray}
and the coefficient $v_{k,n,m}$ is defined as
\begin{align}
v_{k,n,m} &= \frac{1}{2^k \sqrt{(n+m-2k)!}} \times \nonumber \\
& \sum_{u=0}^{\min(k,n,m)} \frac{(-4)^u (n+m-2u)!}{u!(m-u)!(n-u)!(k-u)!}.
\end{align}
\label{thm:heff}
\end{theorem}

\begin{IEEEproof}
See Appendix \ref{app:heff}.
\end{IEEEproof}

The filtered noise samples, $n_{\mathrm{dd}}[k,l]$, are obtained by sampling the continuous filtered noise $n_{\mathrm{dd}}^{w_{\mathrm{rx}}}(\tau,\nu)$ defined in (\ref{discr3}). The statistical correlation between these samples is captured by the noise covariance matrix. The closed-form expression for the elements of this matrix is given by the following theorem.

\begin{theorem}
For the proposed Hermite pulse, the element of the noise covariance matrix corresponding to DD indices $(k_1, l_1)$ and $(k_2, l_2)$ is given by
\begin{align}
&\mathbb{E}[n_{\mathrm{dd}}[k_1,l_1]n_{\mathrm{dd}}^*[k_2,l_2]] = \frac{N_0 \tau_p}{\sigma_\nu} \sum_{q_1,q_2 \in \mathbb{Z}} e^{j2\pi\frac{q_2l_2-q_1l_1}{N}} \nonumber \\
& \times \left( \sum_{n_1,m_1=0}^{N_c-1} c_{2n_1}c_{2m_1} A_{\psi_{2n_1},\psi_{2m_1}}\left(\sigma_\tau (\tau_{k_2,q_2}-\tau_{k_1,q_1}), 0\right) \right) \nonumber \\
& \times \left( \sum_{n_2=0}^{N_c-1} (-1)^{n_2} d_{2n_2} \psi_{2n_2}\left(\frac{\tau_{k_1,q_1}}{\sigma_\nu}\right) \right) \nonumber \\
& \times \left( \sum_{m_2=0}^{N_c-1} (-1)^{m_2} d_{2m_2} \psi_{2m_2}\left(\frac{\tau_{k_2,q_2}}{\sigma_\nu}\right) \right),
\label{eq:noise_cov_closed_form}
\end{align}
where $\tau_{k,q} = \frac{k\tau_p}{M}+q\tau_p$.
\label{thm:noise_cov}
\end{theorem}

\begin{IEEEproof}
See Appendix \ref{app:noise}.
\end{IEEEproof}

\section{Results and Discussions}
\label{sec:results}
\begin{figure}
\hspace{4mm}
\includegraphics[width=8cm,height=5.5cm]{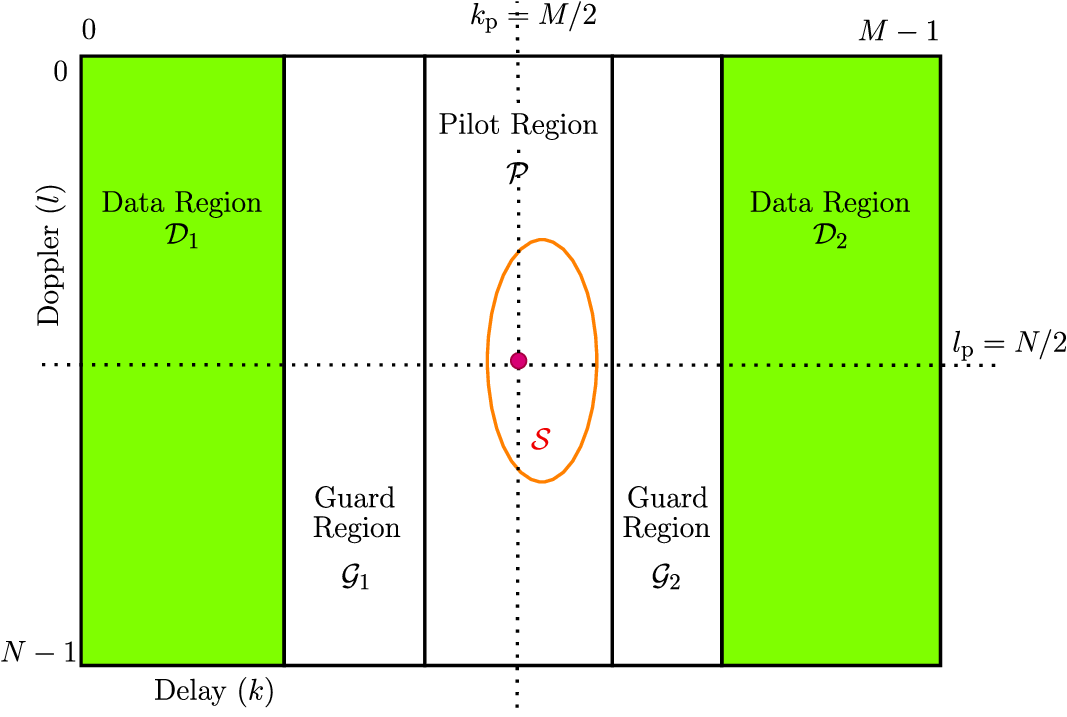}
\caption{Embedded pilot frame with pilot symbol, pilot region ${\mathcal P}$, guard region ${\mathcal G}$, data region ${\mathcal D}$, and support set of the effective channel ${\mathcal S}$.}
\label{fig:embedded_pilot}
\end{figure}

In this section, we evaluate the normalized mean-squared error (NMSE) and BER performance of Zak-OTFS with the proposed Hermite pulse and compare them with those of the sinc, Gaussian, and GS filters. No bandwidth/time expansion ($B'=B, T'=T$) is considered for the pulses. Vehicular-A channel model having $P=6$ paths with fractional DDs and a PDP as detailed in Table \ref{tab_pdp} is used. The maximum Doppler shift is $\nu_{\mathrm{max}}=815$ Hz, and the Doppler shift of the $i$th path is modeled as $\nu_{i}=\nu_{\mathrm{max}}\cos\theta_{i},i=1,\ldots,P$, where $\theta_{i}$s are independent and uniformly distributed in $[0,2\pi)$. Unless stated otherwise, a system configuration with frame size $M=12$ and $N=14$ is used. A Doppler period of $\nu_p = 15$ kHz is considered, with a resulting bandwidth of $B = M\nu_p = 180$ kHz and  frame duration $T = N\tau_p = 0.93$ ms. The considered system parameters satisfy the crystallization condition. Also, in the simulations, the range of values of $m$ and $n$ in (\ref{eqn_channel_matrix}) is limited to -2 to 2, and this is found to ensure an adequate support set of $h_{\mathrm{eff}}[k,l]$ that captures the channel spread accurately. BPSK and 8-QAM modulation alphabets are considered. Model-free I/O relation estimation with an embedded pilot frame \cite{zak_otfs2},\cite{gs}, where a pilot symbol and data symbols are embedded with guard space in between, as shown in Fig. \ref{fig:embedded_pilot}. The pilot symbol is located at $(k_{\mathrm{p}},l_{\mathrm{p}})$ and the pilot region $\mathcal{P}$ spans from $k_{\text{p}}-p_1$ to $k_{\text{p}}+k_{\max}+p_2$, and the guard region $\mathcal{G}$ is defined by the boundaries $k_{\text{p}}-k_{\max}-g_1$ and $k_{\text{p}}+k_{\max}+g_2$. Here, $k_{\max} = \lceil B\max(\tau_{i})  \rceil$ represents the maximum delay spread of the physical channel and $p_1, p_2, g_1, g_2$ are non-negative integers. The additional bins within these regions represented by $p_1, p_2, g_1, g_2$ accommodate the signal spread caused by the pulse shaping filters and can be chosen according to the system bandwidth. The following parameters are used: $(k_{\text{p}},l_{\text{p}})=(M/2,N/2)$, $p_1=p_2=1$, $g_1=1$, $g_2=2$, and $k_{\text{max}}=\lceil B\max(\tau_{i}) \rceil=1 $. A pilot-to-data power ratio (PDR) of 0 dB is used. Minimum mean square error (MMSE) detection is used for recovery of data symbols.

\begin{table}
\centering
\begin{tabular}{|c|c|c|c|c|c|c|}
\hline
Path index ($i$)         & 1 & 2    & 3    & 4    & 5    & 6    \\ \hline
Delay $\tau_{i}$ ($\mu s$)      & 0 & 0.31 & 0.71 & 1.09 & 1.73 & 2.51 \\ \hline
Relative power 
(dB) & 0 & -1   & -9   & -10  & -15  & -20  \\ \hline
\end{tabular}
\caption{Power delay profile of Veh-A channel model.}
\label{tab_pdp}
\vspace{-5mm}
\end{table}

\begin{figure}[t]
\centering \includegraphics[width=0.95\linewidth]{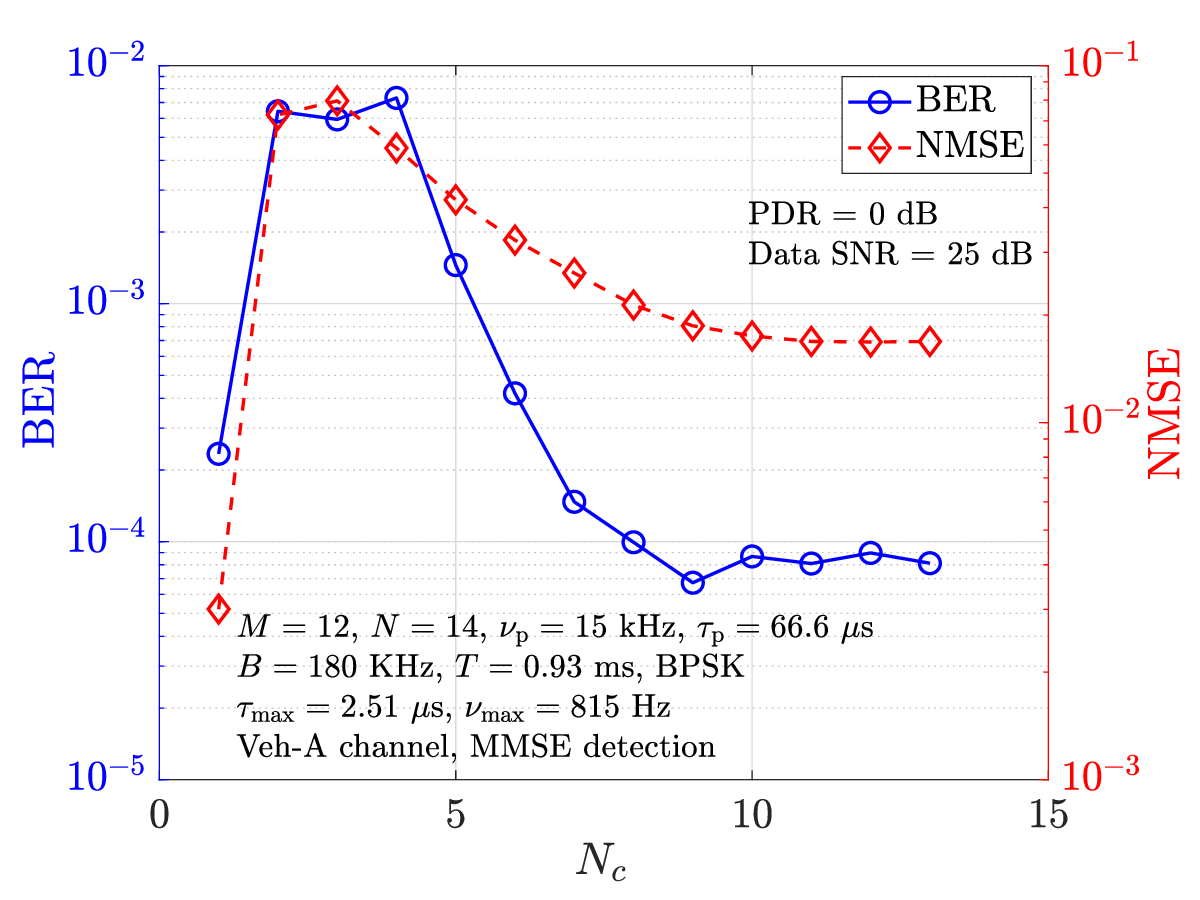}
\vspace{-2mm}
\caption{Effect of the number of basis functions ($N_c$) on the NMSE and BER performance.}
\label{fig:ber_nmse_vs_Nc}
\vspace{-4mm}
\end{figure}

\subsection{Effect of the number of basis functions ($N_c$)}
The key design parameter is $N_c$, the number of even-ordered Hermite basis functions used in the linear combination. This parameter controls the degrees of freedom available for the optimization and governs the fundamental trade-off between orthogonality and localization, as established in Sec. \ref{sec:pulse_chara_analysis}. This, in turn, controls the balance between different sources of interference. To determine the optimal operating point for this trade-off, we evaluate the impact of $N_c$ on the end-to-end system performance. In Fig. \ref{fig:ber_nmse_vs_Nc}, we plot the NMSE 
(defined as the average of $\frac{||\bf{H}_{\mathrm{eff}}-\hat{\bf H}_{\mathrm{eff}}||_{F}^{2}}{||\bf{H}_{\mathrm{}eff}||_{F}^{2}}$, where $\hat{\bf H}_{\mathrm{eff}}$ is the estimated effective channel matrix) and BER as a function of $N_c$ for a fixed data SNR of 25 dB. The results show three distinct regions of behavior as follows.
\begin{enumerate}
\item For small $N_c$ ($< 5$), the optimization is overly constrained, and the resulting pulse has poor orthogonality (high ISI energy). This has a twofold negative impact. First, significant leakage from the data symbols corrupts the pilot, leading to an inaccurate I/O relation estimate and high NMSE. Second, the BER is severely degraded due to both the direct ISI between data symbols and the poor estimate.
\item For intermediate values of $N_c$ ($5 \le N_c \le 10$), performance improves dramatically as the optimization gains enough degrees of freedom to promote orthogonality. The improved null structure reduces data-pilot interference, leading to a steady improvement in the NMSE. The BER curve shows a much more rapid decline (i.e., improved BER). This is because the BER benefits from a twofold improvement: 1) the drastic reduction in direct data-data interference, which is the primary source of detection errors, and 2) the more accurate estimate (lower NMSE), which allows the MMSE detector to operate more effectively. This behavior is a direct reflection of the underlying ISI and sidelobe energy trade-off. As seen in the analysis from Sec. \ref{sec:pulse_chara_analysis}, the ISI energy plummets with increasing $N_c$ while the sidelobe energy rises slowly. The BER is highly sensitive to the ISI, and its steep improvement in this region mirrors the plummeting ISI energy curve, which is the dominant performance factor.
\item For large $N_c$ ($> 10$), the performance saturates. The BER reaches its minimum around $N_c=9$ and then slightly degrades. This indicates that beyond this point, the minimal gains from further ISI reduction are outweighed by the detrimental effects of poorer localization (i.e., higher sidelobe energy).
\end{enumerate}

Based on this analysis, we choose \textbf{$N_c=9$} as a near-optimal choice that strikes a good balance between orthogonality and localization, and 
this value is used in subsequent simulations.

\subsection{Performance comparison with other pulses}
Here, we compare the NMSE and BER performance of the proposed Hermite pulse with those of the sinc, Gaussian, and GS pulses.

\begin{figure}
\centering
\includegraphics[width=0.95\linewidth]{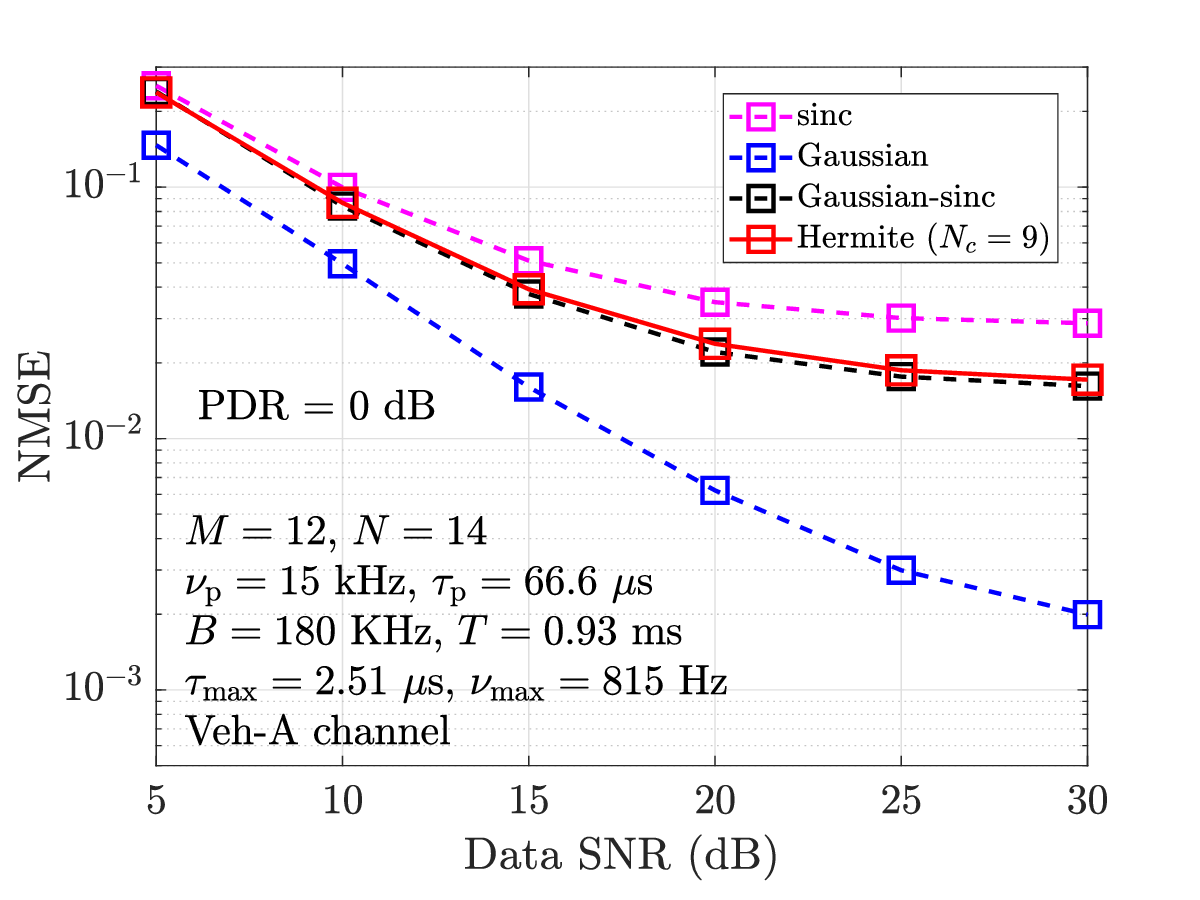}
\caption{NMSE vs data SNR performance comparison between the proposed Hermite pulse and other pulses (sinc, Gaussian, and GS pulses).}
\label{fig:mse_comparison}
\vspace{-4mm}
\end{figure}

{\em NMSE Performance:} Figure \ref{fig:mse_comparison} shows the NMSE performance as a function of data SNR for the Hermite pulse with $N_c=9$ and the sinc, Gauassian, and GS pulses. As established in prior work \cite{gs}, the best-localized Gaussian pulse sets the performance benchmark, achieving the lowest NMSE across all SNRs. In contrast, the sinc, GS, and Hermite pulses exhibit much higher NMSE. This is a direct consequence of their compromised localization manifesting as sidelobes, which introduces two primary sources of estimation error: 1) pilot-data interference, wherein the sidelobe energy from adjacent data symbols leaks into the pilot region used for estimation, and 2) interference from quasi-periodic replicas, wherein the energy leakage from beyond the fundamental period causes DD aliasing. Their combined effect causes the degraded NMSE performance. The sinc pulse, with its high sidelobes, is most susceptible to these effects and, consequently, has the poorest NMSE performance. The proposed Hermite pulse and the GS pulse achieve a much lower NMSE. The fact that their NMSE performance is nearly identical indicates that promoting ISI-free nulls, whether by composite construction (GS) or systematic optimization (proposed Hermite), results in a similar localization penalty relative to the Gaussian ideal.

\begin{figure}
\centering
\includegraphics[width=0.95\linewidth]{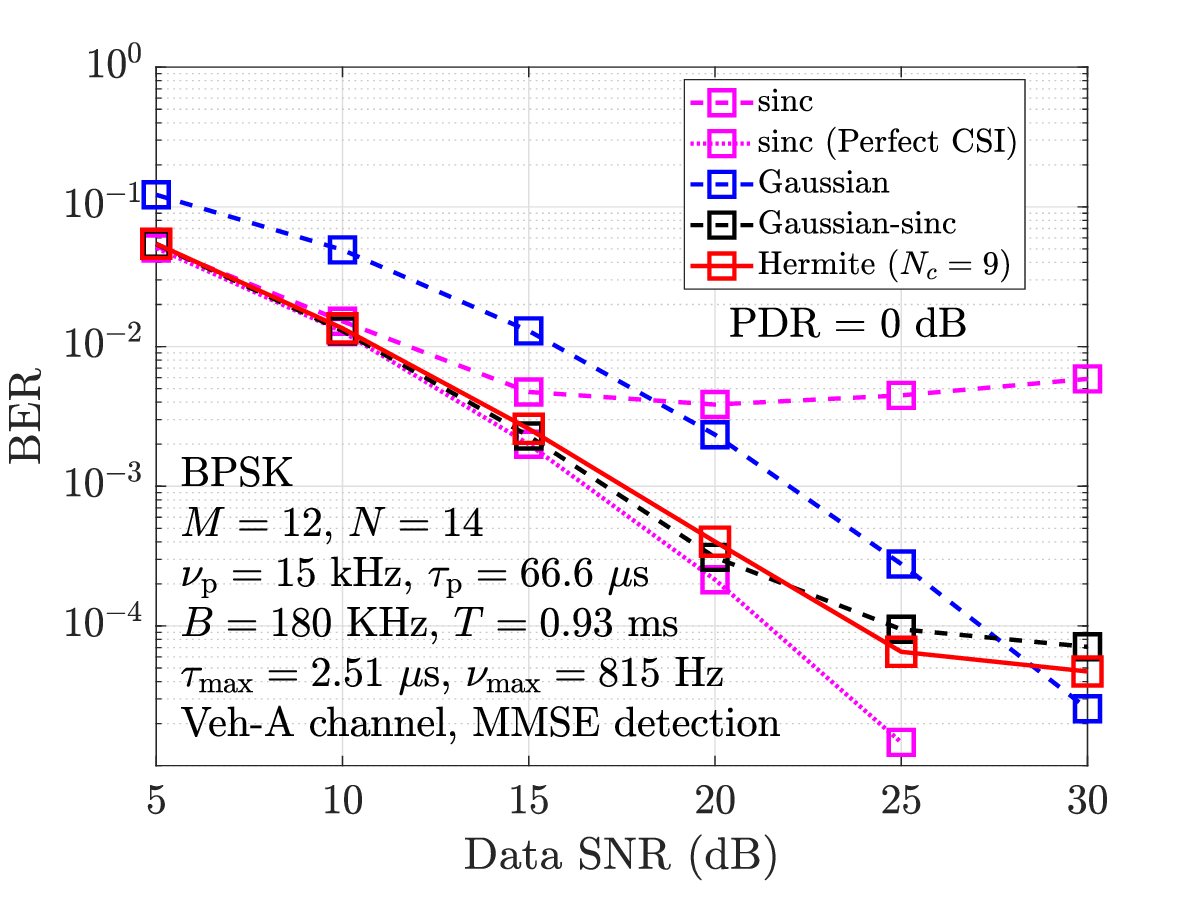}
\vspace{-2mm}
\caption{BER vs data SNR performance comparison between the proposed Hermite pulse and other pulses (sinc, Gaussian, and GS pulses).}
\label{fig:ber_comparison}
\vspace{-4mm}
\end{figure}

\begin{figure}
\centering
\includegraphics[width=0.95\linewidth]{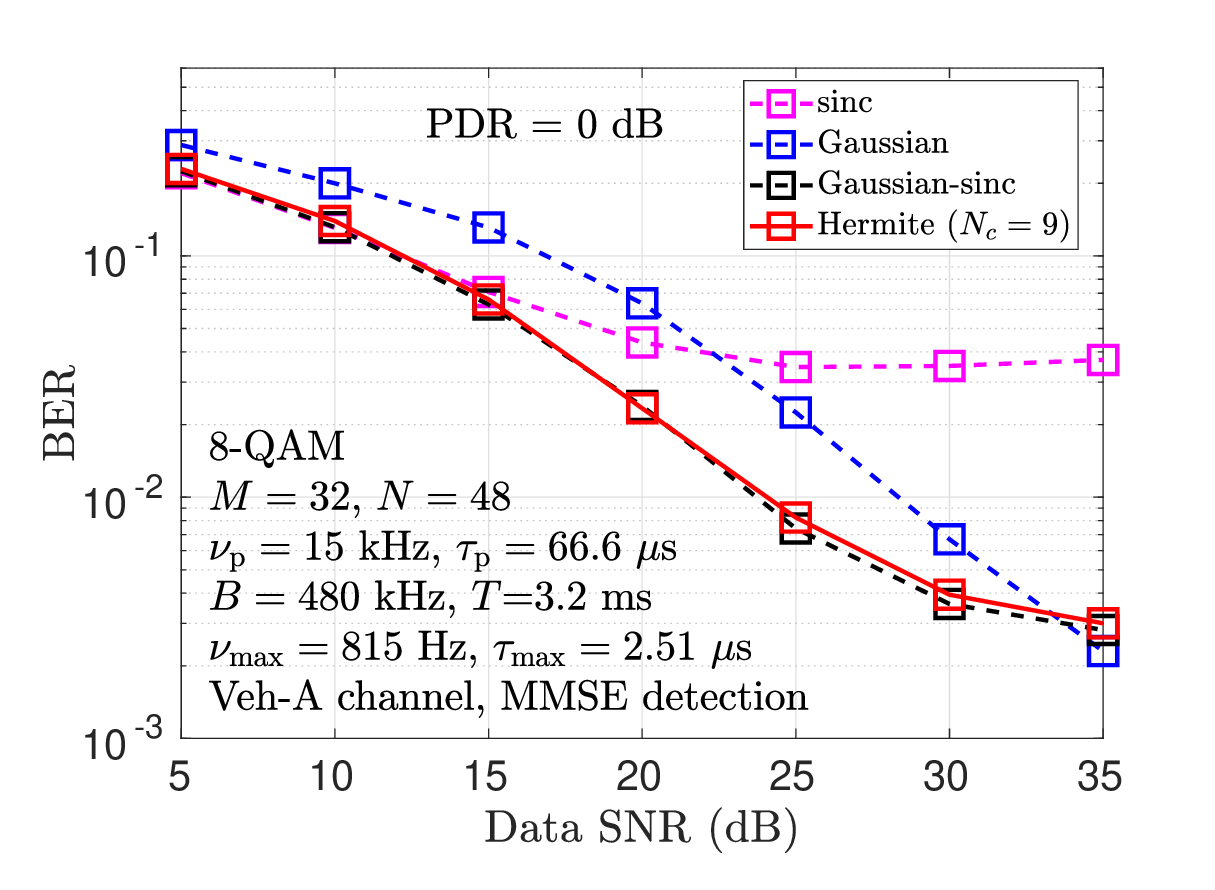}
\vspace{-2mm}
\caption{BER vs data SNR performance comparison between the proposed Hermite pulse and other filters. $M=32, N=48$, 8-QAM.}
\label{fig:ber_comparison_large}
\vspace{-4mm}
\end{figure}

{\em BER performance:} Figure \ref{fig:ber_comparison} shows the achieved BER performance corresponding to the NMSE performance in Fig. \ref{fig:mse_comparison}. At low SNRs, where the noise dominates the interference effects, the Gaussian pulse, despite its excellent estimation accuracy, performs poorly due to overwhelming ISI. The sinc pulse suffers a high error floor at high SNRs, which is a direct result of the poor  NMSE performance. Both the GS pulse and the proposed Hermite pulse ($N_c=9$) successfully navigate this compromise. They significantly outperform the sinc and Gaussian pulses, achieving nearly identical BER performance that is the best among the tested pulses. In Fig. \ref{fig:ber_comparison_large}, a similar BER performance behaviour can be seen for a larger frame size ($M=32,N=48$) and a higher-order modulation alphabet (8-QAM). 

The impressive BER performance achieved by the proposed Hermite pulse is a direct result of a successful design philosophy. Our systematic optimization engineers a pulse that combines two critical properties: 1) the ISI-free null structure of a sinc pulse, which is imitated in the proposed design by construction to ensure robust data detection, and 2) strong localization properties, which, as seen from the NMSE results, lead to a much lower estimation error floor compared to that of the sinc filter. By unifying these two competing requirements, the proposed framework could achieve a state-of-the-art performance level. 

\section{Conclusions}
\label{sec:concl}
Localization and orthogonality attributes of the DD domain pulse in Zak-OTFS influence the I/O relation estimation and data detection performance, respectively. Recognizing the need to strike a fine balance between these two attributes, we developed a systematic framework for designing DD pulses with no bandwidth or time expansion for Zak-OTFS using Hermite basis functions. By constructing the DD pulse as an optimized linear combination of Hermite basis functions, we engineered an optimized pulse that minimized the ISI energy at the DD sampling points while ensuring low sidelobe levels, offering greater design flexibility in terms of control of ISI and sidelobe energies. The optimum coefficients for the linear combination were obtained by solving a constrained optimization problem with ISI energy as the objective to minimize. We derived closed-form expressions of the effective channel and noise covariance, which reduced the simulation run times significantly. Simulation results of Zak-OTFS using embedded pilot and model-free I/O relation estimation in Veh-A channels with fractional DDs showed that the proposed optimized pulses achieve significantly better BER performance compared to that of the canonical sinc and Gaussian pulses and an on-par performance compared to that of the state-of-the-art GS pulse. We note that, among the DD pulses with no bandwidth or time expansion considered in the Zak-OTFS literature so far (sinc, Gaussian, GS, and proposed Hermite pulses), the proposed Hermite pulse and the GS pulse seem to offer the best BER performance with model-free I/O relation estimation. Investigation of other possible pulse designs with further improved localization-orthogonality attributes for Zak-OTFS is open for future research.

\appendices

\section{Proof of Theorem \ref{thm:svd}}
\label{app:svd}
The objective function in (\ref{eq:optimization_problem}) can be written as $\|\mathbf{\Phi}_{\tau} \mathbf{c}\|^2 = \mathbf{c}^T\mathbf{\Phi}_{\tau}^T\mathbf{\Phi}_{\tau}\mathbf{c}$. Using SVD of the real matrix $\mathbf{\Phi}_{\tau} = \mathbf{U}_{\tau}\mathbf{\Sigma}_{\tau}\mathbf{V}_{\tau}^T$, where $\mathbf{U}_{\tau}$ and $\mathbf{V}_{\tau}$ are orthogonal matrices, the objective function becomes
\begin{eqnarray}
\|\mathbf{\Phi}_{\tau} \mathbf{c}\|^2 & \hspace{-2mm} = & \hspace{-2mm} \mathbf{c}^T(\mathbf{U}_{\tau}\mathbf{\Sigma}_{\tau}\mathbf{V}_{\tau}^T)^T(\mathbf{U}_{\tau}\mathbf{\Sigma}_{\tau}\mathbf{V}_{\tau}^T)\mathbf{c} \nonumber \\
& \hspace{-2mm} = & \hspace{-2mm} \mathbf{c}^T\mathbf{V}_{\tau}\mathbf{\Sigma}_{\tau}^2\mathbf{V}_{\tau}^T\mathbf{c} \nonumber \\
& \hspace{-2mm} = & \hspace{-2mm} \mathbf{z}_{\tau}^T\mathbf{\Sigma}_{\tau}^2\mathbf{z}_{\tau}.
\label{norm_exp}
\end{eqnarray}
The last step in (\ref{norm_exp}) follows from a change of variable $\mathbf{z}_{\tau} = \mathbf{V}_{\tau}^T\mathbf{c}$.
Noting that the norm is preserved (i.e., $\|\mathbf{z}_{\tau}\| = \|\mathbf{c}\| = 1$) since $\mathbf{V}_{\tau}$ is orthogonal, the problem can now be restated as
\begin{equation}
\underset{\|\mathbf{z}_{\tau}\|=1}{\text{minimize}} \ \mathbf{z}_{\tau}^T\mathbf{\Sigma}_{\tau}^2\mathbf{z}_{\tau}.
\label{eq:optimization_z}
\end{equation}
The term $\mathbf{z}_{\tau}^T\mathbf{\Sigma}_{\tau}^2\mathbf{z}_{\tau}$ expands to $\sum_{i=1}^{N_c} \sigma_{i,\tau}^2 z_{i,\tau}^2$, where $\sigma_{i,\tau}$s are the singular values of $\mathbf{\Phi}_{\tau}$ in descending order ($\sigma_{1,\tau} \ge \sigma_{2,\tau} \ge \dots \ge \sigma_{N_c,\tau} \ge 0$). This sum is minimized subject to the constraint $\sum z_{i,\tau}^2 = 1$ by selecting $\mathbf{z}_{\tau} = [0 \ 0 \ \dots \ 1]^T$, which places all the energy on the component corresponding to the last singular value, $\sigma_{N_c,\tau}$. The optimal solution is therefore $\mathbf{z}_{\tau, \text{opt}} = [0 \ 0 \ \dots \ 1]^T$, which yields a minimum value of the objective function to be $\sigma_{N_c,\tau}^2$. Expressing the matrix $\mathbf{V}_{\tau}$ in terms of the right singular vectors $\mathbf{v}_{\tau,i}$ as $\mathbf{V}_{\tau} = [\mathbf{v}_{\tau,1} \ \mathbf{v}_{\tau,2} \ \dots \ \mathbf{v}_{\tau,N_c}]$,
the optimal coefficient vector  $\mathbf{c}_{\text{opt}}$ is given by
\begin{equation}
\mathbf{c}_{\text{opt}} = \mathbf{V}_{\tau}\mathbf{z}_{\tau,\text{opt}} 
= \mathbf{v}_{\tau,N_c}
\end{equation}
That is, the optimal coefficient vector $\mathbf{c}_{\text{opt}}$ is the last right singular vector of $\mathbf{\Phi}_{\tau}$.

\section{Proof of Theorem \ref{thm:heff}}
\label{app:heff}
The effective channel, $h_{\mathrm{eff}}(\tau,\nu)$, is given by the twisted convolution cascade of the transmit filter $w_{\mathrm{tx}}(\tau,\nu)$, the physical channel $h_{\mathrm{phy}}(\tau,\nu)$, and the matched receive filter $w_{\mathrm{rx}}(\tau,\nu)$, where $w_{\mathrm{tx}}(\tau,\nu)=w_1({\tau})w_2({\nu})$ and $w_{\mathrm{rx}}(\tau,\nu) = w_{\mathrm{tx}}^*(-\tau,-\nu)e^{j2\pi\nu\tau}$. Therefore, we can write
\vspace{0mm}
\begin{eqnarray}
h_{\mathrm{eff}}(\tau,\nu) & \hspace{-2mm} = & \hspace{-2mm} w_{\mathrm{rx}}(\tau,\nu)*_{\sigma}h_{\mathrm{phy}}(\tau,\nu)*_{\sigma}w_{\mathrm{tx}}(\tau,\nu) \nonumber \\ 
& \hspace{-30mm} = & \hspace{-17mm} w_{\mathrm{rx}}(\tau,\nu)*_{\sigma}\left(\sum_{i=1}^{P}h_{i}\delta(\tau-\tau_{i})\delta(\nu-\nu_{i})\right)*_{\sigma}w_{1}(\tau)w_{2}(\nu) \nonumber    \\ 
& \hspace{-30mm} = & \hspace{-17mm} w_{1}^{*}(-\tau)w_{2}^{*}(-\nu)e^{j2\pi\nu\tau}*_{\sigma}\bigg(\sum_{i=1}^{P}h_{i}w_{1}(\tau-\tau_{i})w_{2}(\nu-\nu_{i}) \nonumber \\ 
& \hspace{-30mm} & \hspace{-17mm} e^{j2\pi\nu_{i}(\tau-\tau_{i})}\bigg) \nonumber \\
& \hspace{-30mm} = & \hspace{-17mm} \sum_{i=1}^{P}
\left(\int w_{1}^{*}(-\tau')w_{1}(\tau-\tau_{i}-\tau') e^{-j2\pi\nu_{i}\tau'}d\tau'\right) \nonumber \\
& \hspace{-30mm} & \hspace{-17mm} 
\left(\int \hspace{-1mm} w_{2}^{*}(-\nu')w_{2}(\nu-\nu_{i}-\nu')e^{j2\pi\nu'\tau}d\nu'\hspace{-0.5mm} \right) \hspace{-0.5mm} h_{i}e^{j2\pi\nu_{i}(\tau-\tau_{i})} \nonumber \\
& \hspace{-30mm} = & \hspace{-17mm}
\sum_{i=1}^{P} h_i e^{j2\pi\nu_i(\tau-\tau_i)} \mathcal{K}_i^{w_1}(\tau) \mathcal{K}_i^{w_2}(\tau,\nu),
\label{eqn:channel_matched}
\end{eqnarray}
where the two integral kernels $\mathcal{K}_i^{w_1}(\tau)$ and $\mathcal{K}_i^{w_2}(\tau,\nu)$ correspond to the delay and Doppler pulses, respectively, 
given by
\begin{eqnarray}
\hspace{-4mm}
\mathcal{K}_i^{w_1}(\tau) & \hspace{-2mm} = & \hspace{-2mm} \int w_1^*(-\tau')w_1(\tau-\tau_i-\tau')e^{-j2\pi\nu_i\tau'}d\tau', \\
\hspace{-4mm}
\mathcal{K}_i^{w_2}(\tau,\nu) & \hspace{-2mm} = & \hspace{-2mm} \int w_2^*(-\nu')w_2(\nu-\nu_i-\nu')e^{j2\pi\nu'\tau}d\nu'.
\end{eqnarray}
These kernels can be related to the symmetric ambiguity function. The symmetric cross-ambiguity function between two functions $f(t)$ and $g(t)$ is defined as \cite{herm_amb}
\begin{equation}
A_{f,g}(\tau, \nu) \triangleq \int f(t+\tau/2)g^*(t-\tau/2)e^{j2\pi\nu t} dt.
\label{amb_def}
\end{equation}
By performing a change of variables, the kernels can be related to the auto-ambiguity functions of $w_1$ and $w_2$ as
\begin{eqnarray}
\mathcal{K}_i^{w_1}(\tau) & \hspace{-2mm} = & \hspace{-2mm} e^{-j\pi\nu_i(\tau-\tau_i)} A_{w_1,w_1}(\tau_i-\tau, -\nu_i), \label{kernel1} \\
\mathcal{K}_i^{w_2}(\tau,\nu) & \hspace{-2mm} = & \hspace{-2mm} e^{j\pi\tau(\nu-\nu_i)} A_{w_2,w_2}(\nu_i-\nu, \tau). \label{kernel2}
\end{eqnarray}
We use the definition of ambiguity function in (\ref{amb_def}), expand $w_1(\tau)$ using (\ref{eq:w1_pulse}), and use the linearity property to express the auto-ambiguity function of the pulse as a double summation of the cross-ambiguity functions of its constituent basis functions to write
\begin{equation}
A_{w_1,w_1}(\tau, \nu) = \sum_{n_1,m_1=0}^{N_c-1} c_{2n_1}c_{2m_1} A^{(\tau)}_{\phi_{2n_1},\phi_{2m_1}}(\tau, \nu),
\end{equation}
where $A^{(\tau)}_{\phi_{n},\phi_{m}}(\cdot,\cdot)$ denotes the cross-ambiguity function between the scaled Hermite basis functions parameterized by $\sigma_{\tau}$.

The basis functions $\phi_n(\sigma,t) = \sqrt{\sigma}\psi_n(\sigma t)$ are scaled versions of the unscaled basis functions $\psi_n(t)$. The ambiguity function of such scaled functions has a known scaling property. For two functions $\tilde{f}(t)=\sqrt{\sigma}f(\sigma t)$ and $\tilde{g}(t)=\sqrt{\sigma}g(\sigma t)$, the ambiguity function is given by
\begin{equation}
A_{\tilde{f},\tilde{g}}(\tau,\nu) = A_{f,g}(\sigma\tau, \nu/\sigma).
\end{equation}
Applying this property to our basis functions, we get $A^{(\tau)}_{\phi_{n},\phi_{m}}(\tau, \nu) = A_{\psi_n,\psi_m}(\sigma_\tau\tau, \nu/\sigma_\tau)$. This leads to the expansion for the ambiguity function of the pulse $w_1$ as 
\begin{equation}
A_{w_1,w_1}(\tau', \nu') = \hspace{-2mm} \sum_{n_1,m_1=0}^{N_c-1} \hspace{-2mm} c_{2n_1}c_{2m_1} A_{\psi_{2n_1},\psi_{2m_1}}\left(\sigma_\tau\tau', \frac{\nu'}{\sigma_\tau}\right).
\label{exp1}
\end{equation}
A similar expansion applies to $A_{w_2,w_2}(\nu'', \tau'')$ that yields
\begin{equation}
A_{w_2,w_2}(\nu'', \tau'') = \hspace{-2mm} \sum_{n_2,m_2=0}^{N_c-1} \hspace{-2mm} d_{2n_2}d_{2m_2} A_{\psi_{2n_2},\psi_{2m_2}}\left(\sigma_\nu\nu'', \frac{\tau''}{\sigma_\nu}\right).
\label{exp2}
\end{equation} 
Substituting the expansions in (\ref{exp1}) and (\ref{exp2}) into the expressions for the kernels in (\ref{kernel1}) and (\ref{kernel2}), which when used in (\ref{eqn:channel_matched}) yields the expression (\ref{eq:heff_closed_form}) in Theorem \ref{thm:heff}.

\section{Proof of Theorem \ref{thm:noise_cov}}
\label{app:noise}
We begin with the definition of the filtered noise in the continuous DD domain, which is given by the twisted convolution of the receive filter $w_{\mathrm{rx}}(\tau,\nu)$ and the DD domain noise $n_{\mathrm{dd}}(\tau,\nu)$, i.e.,
\begin{equation}
n_{\mathrm{dd}}^{w_{\mathrm{rx}}}(\tau,\nu) = w_{\mathrm{rx}}(\tau,\nu) *_\sigma n_{\mathrm{dd}}(\tau,\nu).
\end{equation}
Using the definition of the matched receive filter $w_{\mathrm{rx}}(\tau,\nu)$ in the definition of twisted convolution, the expression becomes:
\begin{eqnarray}
n_{\mathrm{dd}}^{w_{\mathrm{rx}}}(\tau,\nu) & \hspace{-2.5mm} = & \hspace{-4mm} \iint \hspace{-1mm} w_1^*(-\tau')w_2^*(-\nu')e^{j2\pi\nu'\tau'} n_{\mathrm{dd}}(\tau-\tau',\nu-\nu') \nonumber \\
& \hspace{-2.5mm} & \hspace{-4mm} \times e^{j2\pi\nu'(\tau-\tau')} d\tau'd\nu'.
\end{eqnarray}
Substituting the Zak transform of $n(t)$,  $n_{\mathrm{dd}}(\tau,\nu) = \sqrt{\tau_p} \sum_{q\in\mathbb{Z}} n(\tau+q\tau_p) e^{-j2\pi\nu q\tau_p}$, and separating the integrals, this can be written as a sum over the index $q$ from the Zak transform of the noise as
\begin{equation}
n_{\mathrm{dd}}^{w_{\mathrm{rx}}}(\tau,\nu) = \sqrt{\tau_p} \sum_{q\in\mathbb{Z}} e^{-j2\pi\nu q \tau_p} \mathcal{J}_{q,n}^{w_1}(\tau) \mathcal{J}_{q}^{w_2}(\tau),
\label{eq:ndd_kernels}
\end{equation}
where the two integral kernels are
\begin{eqnarray}
\mathcal{J}_{q,n}^{w_1}(\tau) & \hspace{-2mm} = & \hspace{-2mm} \int w_1^*(-\tau') n(\tau-\tau'+q\tau_p) d\tau', \\
\mathcal{J}_{q}^{w_2}(\tau) & \hspace{-2mm} = & \hspace{-2mm} \int w_2^*(-\nu') e^{j2\pi\nu'(\tau+q\tau_p)} d\nu'.
\end{eqnarray}
The discrete noise samples $n_{\mathrm{dd}}[k,l]$s are obtained by sampling $n_{\mathrm{dd}}^{w_{\mathrm{rx}}}(\tau,\nu)$ at the DD grid points $(\tau, \nu) = (\frac{k\tau_p}{M}, \frac{l\nu_p}{N})$. Using (\ref{eq:ndd_kernels}) and sampling, the noise covariance can be expressed as 
\begin{eqnarray}
\mathbb{E}[n_{\mathrm{dd}}[k_1,l_1]n_{\mathrm{dd}}^*[k_2,l_2]] & \hspace{-2.5mm} = & \hspace{-2.5mm} \tau_p \hspace{-1mm} \sum_{q_1,q_2 \in \mathbb{Z}} \hspace{-2mm}  e^{j2\pi\frac{q_2l_2-q_1l_1}{N}} \mathcal{J}_{q_1}^{w_2}\hspace{-0.5mm}\left(\frac{k_1\tau_p}{M}\right) \nonumber \\ 
& \hspace{-50mm} & \hspace{-40mm} \times 
\mathcal{J}_{q_2}^{w_2*}\left(\frac{k_2\tau_p}{M}\right) \mathbb{E}\left[\mathcal{J}_{q_1,n}^{w_1}\left(\frac{k_1\tau_p}{M}\right) \mathcal{J}_{q_2,n}^{w_1*}\left(\frac{k_2\tau_p}{M}\right)\right].
\label{eq:app_Rn_start}
\end{eqnarray}
Next, using the fact that $w_2(t)$ is real and even, we solve for the deterministic kernel $\mathcal{J}_{q}^{w_2}(\tau)$ as
\begin{align}
\mathcal{J}_{q}^{w_2}(\tau) &= \int w_2^*(-\nu') e^{j2\pi\nu'(\tau+q\tau_p)} d\nu' \nonumber \\
& = \mathcal{F}\{w_2(t)\}^*(\tau+q\tau_p)) \nonumber \\
& = \sum_{m=0}^{N_c-1} d_{2m} \mathcal{F}\{\phi_{2m}(\sigma_\nu, t)\}(\tau+q\tau_p)) \nonumber \\
& = \frac{1}{\sqrt{\sigma_\nu}} \sum_{m=0}^{N_c-1} (-1)^{m} d_{2m} \psi_{2m}\left(\frac{\tau+q\tau_p}{\sigma_\nu}\right).
\label{eq:app_J_w2_result}
\end{align}
The final step uses the Fourier transform of the scaled Hermite basis function (derived below) and simplifies using $(-j)^{2m} = (-1)^m$.

The derivation for the scaled Hermite function $\phi_k(\sigma,t)$ uses the fact that the standard Hermite function $\psi_k(t)$ is an eigenfunction of the Fourier transform operator \cite{herm_ft}:
\begin{equation}
\mathcal{F}\{\psi_k(t)\}(f) = (-j)^k \psi_k(f).
\label{eq:hermite_eigen_property}
\end{equation}
Substituting this property, we write
\begin{eqnarray}
\mathcal{F}\{\phi_k(\sigma, t)\} (f) & \hspace{-2mm} = & \hspace{-2mm} \mathcal{F}\{\sqrt{\sigma} \psi_k(\sigma t)\} (f) \quad \ \text{(by definition)} \nonumber \\
& \hspace{-37mm} = & \hspace{-20mm} \sqrt{\sigma} \left[ \frac{1}{\sigma} \mathcal{F}\{\psi_k(t)\} \left(\frac{f}{\sigma}\right)\right] \quad \text{(by scaling property)} \nonumber \\
& \hspace{-37mm} = & \hspace{-20mm} \frac{\sqrt{\sigma}}{\sigma} \left[ (-j)^k \psi_k \left(\frac{f}{\sigma}\right) \right] \quad \quad \quad \quad \quad \ \ \text{(using \eqref{eq:hermite_eigen_property})} \nonumber \\
& \hspace{-37mm} = & \hspace{-20mm} \frac{1}{\sqrt{\sigma}} (-j)^k \psi_k\left(\frac{f}{\sigma}\right).
\end{eqnarray}
Next, solving the expectation over the stochastic kernels using the property of white noise, $\mathbb{E}[n(t_1)n^*(t_2)]=N_0\delta(t_1-t_2)$, we write
\begin{eqnarray}   \mathbb{E}\left[\mathcal{J}_{q_1,n}^{w_1}\left(\frac{k_1\tau_p}{M}\right) \mathcal{J}_{q_2,n}^{w_1*}\left(\frac{k_2\tau_p}{M}\right)\right] & \hspace{-2mm} = & \hspace{-2mm} \iint w_1^*(-\tau_1')w_1(-\tau_2') \nonumber \\ 
& \hspace{-60mm} & \hspace{-50mm} \underbrace{\mathbb{E}[n(\frac{k_1\tau_p}{M}+q_1\tau_p-\tau_1')n^*(\frac{k_2\tau_p}{M}+q_2\tau_p-\tau_2')]}_{=N_{0}\delta\left(\tau_{2}-\tau_{1}-\frac{(k_{2}-k_{1})\tau_{p}}{M}-(q_{2}-q_{1})\tau_{p}\right)} d\tau_1'd\tau_2' \nonumber \\
& \hspace{-102mm} = & \hspace{-52mm} N_0 \int w_1(\tau_1')w_1(\tau_1' - (\tau_{k_1,q_1}-\tau_{k_2,q_2})) d\tau_1' \nonumber \\
& \hspace{-102mm} = & \hspace{-52mm} N_0  A_{w_1,w_1}(\tau_{k_2,q_2}-\tau_{k_1,q_1}, 0) \nonumber \\
& \hspace{-102mm} = & \hspace{-52mm} N_0 \hspace{-2mm} \sum_{n_1,m_1=0}^{N_c-1} \hspace{-3mm} c_{2n_1}c_{2m_1} A_{\psi_{2n_1},\psi_{2m_1}}\hspace{-1mm}\left(\sigma_\tau (\tau_{k_2,q_2}-\tau_{k_1,q_1}), 0\right).
\label{eq:app_J_w1_result}
\end{eqnarray}
Finally, substituting (\ref{eq:app_J_w2_result}) and (\ref{eq:app_J_w1_result}) in (\ref{eq:app_Rn_start}), we arrive at the result presented in Theorem \ref{thm:noise_cov}.


\end{document}